\documentclass[useAMS,usenatbib]{mn2e}
\usepackage{amssymb,amsmath,graphicx}

\def\A{$\mathcal{A}$}
\def\B{$\mathcal{B}$}
\def\C{$\mathcal{C}$}
\def\D{$\mathcal{D}$}

\def\ctn{$CTN$}

\def\lya{Ly$\alpha$}
\def\halpha{H$\alpha$}
\def\hbeta{H$\beta$}

\def\ebv{$E(B-V)$}
\def\ebvs{$E(B-V)_\star$}
\def\ebvg{$E(B-V)_\mathrm{IS}$}
\def\bet{$\beta$}

\def\flya{$F_{\mathrm{Ly}\alpha}$}
\def\wlya{$W_{\mathrm{Ly}\alpha}$}

\def\flyc{$f_{900}$}
\def\ffuv{$f_{1500}$}
\def\fnuv{$f_{2200}$}

\def\hi{H{\sc i}}

\def\nii{N{\sc ii}}

\def\nhi{$n_\mathrm{HI}$}

\def\aap{A\&A}
\def\aaps{A\&AS}
\def\aj{AJ}
\def\apj{ApJ}
\def\apjs{ApJS}
\def\apjl{ApJL}
\def\araa{ARA\&A}
\def\mnras{MNRAS}
\def\pasp{PASP}
\def\pasj{PASJ}
\def\procspie{Proc. SPIE}

\title[The escape of Lyman photons from a young starburst]
{The escape of Lyman photons from a young starburst: \\
the case of Haro\,11
\thanks{Based on observations 
made with the NASA/ESA Hubble Space Telescope, obtained  at the
Space Telescope Science Institute, which is operated by the Association
of Universities for Research in Astronomy, Inc., under NASA contract NAS
5-26555. These observations are associated with programs \#GO\,9470 and 
\#GO\,10575.}
\thanks{Based on observations made with ESO Telescopes at the La Silla 
Observatories under programme ID 073.B-0785.}
}

\author[M. Hayes et al.]{Matthew Hayes$^{1}$\thanks{E-mail:
matthew@astro.su.se},
G\"oran \"Ostlin$^{1}$, 
Hakim Atek$^{2}$, 
Daniel Kunth$^{2}$, \newauthor
J. Miguel Mas-Hesse$^{3}$, 
Claus Leitherer$^{4}$,
Elena Jim{\'e}nez-Bail{\'o}n$^{5}$, 
and  
Angela Adamo$^{1}$
\\
$^{1}$Stockholm Observatory, AlbaNova University Centre, 
106 91 Stockholm, Sweden \\
$^{2}$Institut d'Astrophysique de Paris, Paris (IAP), 98 bis boulevard Arago, 
75014 Paris, France \\
$^{3}$Centro de Astrobiolog\'{\i}a (CSIC-INTA), E28850 Torrejon de Ardoz, 
Madrid, Spain\\
$^{4}$Space Telescope Science Institute, 3700 San Martin Drive, Baltimore, 
MD 21218, USA\\
$^{5}$Laboratorio de Astrof\'{\i}sica Espacial y F\'{\i}sica Fundamental
(LAEFF--INTA), POB 78, 28691 Villanueva de la Ca\~nada, Spain }

\begin{document}

\date{Accepted 2007 September 17. Received 2007 September 12; in original form
2007 July 05}

\pagerange{\pageref{firstpage}--\pageref{lastpage}} \pubyear{2007}

\maketitle

\label{firstpage}

\begin{abstract}

Lyman-alpha (\lya) is one of the dominant tools used to probe the
star-forming galaxy population at high-redshift ($z$). 
However, astrophysical interpretations of data drawn from \lya\ alone hinge on
the \lya\ escape fraction which, due to the complex radiative transport, 
may vary greatly.
Here we map the \lya\ emission from the local luminous blue 
compact galaxy Haro\,11, a known emitter of \lya\ and the only known candidate 
for low-$z$ Lyman continuum emission (LyC). 
To aid in the interpretation we perform a detailed UV and optical multi-wavelength 
analysis and model the stellar population, dust distribution, ionising photon 
budget, and star-cluster population. 
We use archival X-ray observations to further constrain properties of the
starburst and estimate the neutral hydrogen column density. 

The \lya\ morphology is found to be largely symmetric around a single young 
star forming knot and is strongly decoupled from other wavelengths.
From general surface photometry, 
only very slight correlation is found between \lya\ and \halpha, \ebv, and the 
age of the stellar population. 
Only around the central \lya-bright cluster do we find the \lya/\halpha\ ratio 
at values predicted by recombination theory. 
The total \lya\ escape fraction is found to be just 3\%.
We compute that $\sim 90$\% of the \lya\ photons that escape do so after
undergoing multiple resonance scattering events, masking their point of origin.
This leads to a largely symmetric distribution and, by increasing the distance
that photons must travel to escape, decreases the escape probability significantly.
While dust must ultimately be responsible for the destruction of \lya, it plays
little role in governing the observed morphology, which is regulated more by ISM
kinematics and geometry. 
We find tentative evidence for local \lya\ equivalent width in the immediate 
vicinity of star-clusters being a function of cluster age, consistent with 
hydrodynamic studies. 
We estimate the intrinsic production of ionising photons and put
further constraints of $\sim 9$\% on the escaping fraction of photons at 
900\AA.

\end{abstract}


\section{Introduction}

Four decades ago
\cite{pp67} 
discussed the prospects of identifying `primeval' galaxies (i.e. galaxies
forming their first generation of stars), using both the Lyman decrement and
Lyman-alpha (\lya) emission line as observational tracers. 
Both methods rely upon the galaxies hosting numerous young, massive
stars producing a strong radiation field in the far ultraviolet (UV). 
The absorption of this radiation field bluewards of the Lyman absorption edge
results in the Lyman-break phenomenon, while the reprocessing of the absorbed
photons in astrophysical nebulae results in the superimposition of strong 
hydrogen recombination lines on the galaxy spectra. 
While it's likely that both features are present in the spectra of high-redshift
($z$) starbursts, from a survey perspective they compete.
Lyman-break candidates are identified in multi-broadband imaging surveys where
large ranges in redshift (and therefore cosmic volume) can be probed, but are
biased towards the detection of 
galaxies with FUV continua towards the brighter end of the luminosity function 
(LF) -- less numerous in the hierarchical galaxy formation scenario. 
On the other hand, because emission lines concentrate a large amount of energy
in a very small spectral region, sources with much fainter continuum can be
uncovered. Unfortunately, isolation of a line requires either a spectroscopic
surveys which typically probe narrow fields of view because of the slit
dramatically reducing the size in one dimension, or 
narrowband imaging which probes only a small range in redshift.  
The advent of large diameter reflectors, efficient optics, and sensitive
photometric devices has resulted in both techniques enjoying much success in
recent years.

Young galaxies with low dust contents can be expected to produce \lya\ emission
with high equivalent widths (\wlya), $\sim 240$\AA\ for solar-like metallicities
($Z$) and standard initial mass functions (IMF) 
\citep{charlotfall93}. 
This value is predicted to increase significantly as metallicities approach 
the population {\sc iii} domain 
\citep{schaerer03}.
Since the \lya\ rest wavelength lies in the FUV, the line is still observable
from the ground in the optical domain at $z\sim 6$ and beyond making it, in
principle, the ideal tool by which to identify star-forming galaxies in the early
universe. 
Despite a rocky start 
\citep{pritchet93}
\lya\ emitters (LAEs) are showing up en masse in high-$z$ surveys and
perhaps their observed and predicted number-densities are now converging 
\citep{ledelliou06}. 
\lya\ is now being used to put constraints on the final stages of cosmic
reionisation 
\citep{malhotrarhoads04,dijkstra06} and 
explore high-redshift large scale structure and galaxy clustering
\citep{hamana04,malhotra05,murayama07}. 
In addition, numerous sources have been detected through narrowband imaging
techniques with very high \wlya
\citep{malhotrarhoads02,shimasaku06}, 
perhaps indicative of top heavy IMFs or extreme
stars, making such targets ideal to search for signatures of the so-far 
illusive population {\sc iii} stars. 
However, so far spectroscopic observations have failed to detect the strong
He{\sc ii}$\lambda 1640$\AA\ feature expected from pop {\sc iii} objects 
(eg. \citealt{dawson04,nagao07}).
Finally, \lya\ is also being used to estimate the cosmic star formation rate 
density at the highest redshifts
\citep{fujita03,yamada05}. 
Such star formation rates (SFR) are typically estimated assuming case B
recombination 
\citep{brocklehurst71} 
and the \halpha\ SFR calibration of 
\cite{kennicutt98}, 
although the large spread in SFR(\lya) {\em vs.} SFR(FUV) and consistent
underestimates from SFR(\lya)
(eg. \citealt{murayama07})
suggest that such a calibration should be used with caution. 
This caution must be extended to all cosmological studies in light of the fact
that only a fraction of UV-selected high-$z$ targets show a \lya\
feature in emission 
\citep[e.g.][]{shapley03}. 

The complexities of using \lya\ as a cosmological tool result from the fact that
it is a resonant line, and its potential cosmological importance has motivated a
number of studies of star-forming galaxies at low-$z$ and theoretical models of
resonant line radiative transfer. 
Early studies with the {\em International Ultraviolet Explorer (IUE)} initially
mirrored the first results at high-$z$: \lya\ was typically weak or absent in local
starbursts. 
This systematic weakening of \lya\ was first attributed to dust absorption 
(eg. \citealt{charlotfall91}) 
and an anticorrelation between \wlya\ and metallicity was found in the {\em IUE}
sample
\citep{charlotfall93}. 
However, the damped \lya\ absorption seen in some local starbursts, and the
failure of dust corrections to reconcile \lya\ with the fluxes predicted by
recombination theory
\citep{giavalisco96}, 
is indicative of selective attenuation of \lya.
This is clearly exemplified by the fact that the most metal-poor galaxies known
at low-$z$ ({\sc i}Zw\,18 and SBS\,03350-52) are shown to be damped \lya\ absorbers 
\citep{kunth94,thuanizotov97} in spectroscopic observations from the {\em
Goddard High Resolution Spectrograph (GHRS)} and {\em Space Telescope
Imaging Spectrograph (STIS)}. 
This is to be expected if the starburst is enshrouded by a static layer of
neutral hydrogen that is able to resonantly trap \lya, thereby greatly 
increasing the path-lengths of the photons in order to escape the host,
and exponentially increasing the chance of their destruction by dust
grains 
\citep{neufeld90}.
Further {\em GHRS} observations 
\citep{kunth98} 
showed that when \lya\ is seen in emission, it almost systematically
shows a P\,Cygni profile and systematic velocity offset from low ionisation
state metal absorption features in the neutral ISM, suggesting that an
outflowing medium is an essential ingredient in the formation of the line 
profile and \lya\ escape physics. 
Similar results have also been found at high-$z$
\citep{shapley03,tapken07}.
Hydrodynamic models of expanding bubbles 
\citep{tt99} 
predicted \lya\ line profiles to follow an evolutionary sequence starting with
pure absorption at the earliest times, developing through a pure emission phase
into P\,Cygni profiles as the ISM is driven out by mechanical feedback from the
starburst, and fading back into absorption at late times.
This allowed 
\cite{mashesse03} 
to reconcile the variety of \lya\ profiles observed at low-$z$
in such an evolutionary sequence. 
Further theoretical studies
\citep[e.g.][]{ahn03,verhamme06} 
have shown how a wide variety of P\,Cygni-like and asymmetric profiles can
develop, depending on the ISM properties and geometry. 
The line formation becomes more complex still if the ISM is multiphase 
\citep{neufeld91,hansen06}
and certain physical and kinematic configurations may lead to increases in the
fraction of escaping \lya\ photons compared to non-resonant radiation, thereby
effectively boosting \wlya. 
Recent advances have been made in the understanding of observed line profiles
with the implementation of full 3-D codes with arbitrary distributions of gas,
ionisation, temperature, dust, and kinematics 
\citep{verhamme06}. 

Since nebular ionisation does not occur in situ with ionising sources, 
recombination line imaging (e.g. \halpha) may reveal a morphology different from
that found by imaging of the stellar continuum.
This phenomenon may be much more significant for \lya\ photons which 
resonantly scatter.
While \lya\ and \halpha\ have the same points of origin, it is likely that 
internal \lya\ scattering events may cause \lya\ photons to be emitted far from 
their production sites, and not be spatially correlated with \halpha\ or other
non-resonant recombination lines. 
Targeted spectroscopic observations are therefore liable to miss a fraction of
the emission and, while containing neither frequency nor kinematic information,
\lya\ imaging becomes an invaluable complement to the spectroscopic studies. 
This was the motivation for our imaging survey of local starbursts using the
{\em Advanced Camera for Surveys (ACS)}
(\citealt{kunth03,hayes05,ostlin07} in preparation). 
This is a truly unique dataset for a number of reasons. 
Firstly the angular resolution of the {\em ACS} allows us to map the \lya\
morphology on scales of 5--15~pc; 2--3 orders of magnitude better than 
typical studies at high-$z$. 
The addition of \halpha\ (absent in almost all high-$z$ studies due to an 
inconvenient rest wavelength) allows us to quantitatively examine the decoupling 
of \lya\ from non-resonant lines and estimate the global escape fractions.
Multiband UV and optical data allow us to map dust reddening, stellar ages and
masses, ionising photon production, and other properties of the host,
all at the same resolution as \lya.
The first detailed \lya\ imaging study of a local starburst, ESO\,338-IG04
\citep{hayes05} 
found emission and absorption varying on very small scales in the
central starburst regions,  and little or no correlation with the FUV morphology. 
The starburst is surrounded with a large, diffuse, low surface brightness \lya\
halo that contributes $\sim 70$\% to the global \lya\ luminosity, resulting from
the resonant decoupling and diffusion of \lya. 
The total escape fraction  was found to be just $\sim 5$\%, implying any global
values (eg. SFR) that would be estimated from \lya\ alone would be seriously at
fault.

Feedback processes from star-formation are capable of driving galaxy-scale
`superwinds' that shock heat and accelerate the ambient medium and
circumnuclear gas, resulting in large-scale, diffuse X-ray nebulae
\citep{heckman90}. 
Typically, hot X-ray emitting regions exhibit a tight morphological correlation 
with with warm gas as probed by optical emission lines
\citep{grimes05}. 
Indeed, such X-ray nebulae are near-ubiquitous in starbursts 
\citep{strickland04} 
and have been calibrated as tracers of star-formation rates
\citep{ranalli03,colbert04}.
\cite{grimes05} 
have also shown that observed X-ray spectra are well reproduced by a single 
thermal plasma  over a range of galaxy classifications spanning 
dwarfs, discs, and ultra-luminous infrared galaxies (ULIRG), although the central
regions of some ULIRGs require an additional power-law component. 
Since the diffuse X-ray component can be so
valuable for an understanding of the wind properties, and feedback is 
an essential ingredient in the escape of \lya\ 
\citep{kunth98}, 
X-ray information provides a valuable supporting dataset for a detailed
investigation of \lya.

In this article we turn our attention to another target in our sample, the 
well-known, luminous ($M_B$ = -20.5), low metallicity ($\log(O/H)+12=7.9$;
\citealt{bergvall02}) blue compact galaxy (BCG).
It exhibits a complex morphology consisting of three main star-forming 
condensations and an unrelaxed kinematic structure
\citep{ostlin99,ostlin01}, 
suggestive of a dwarf galaxy merger. 
It is actively star-forming, exhibits a large number of bright young star
clusters 
\citep{ostlin00},
is a known emitter of \lya\ 
\citep{kunth98}, 
and the only known local candidate emitter of Lyman continuum (LyC) 
\citep{bergvall06,grimes07}. 
The FUV continuum luminosity puts Haro\,11 at the brighter end of the
distribution of \lya\ emitters at $z\sim 3.1$
\citep{gronwall07}.
We utilise {\em HST/ACS} images in the FUV ({\em Solar Blind Channel; SBC}), to
examine \lya\ and nearby continuum, and 
broadband images in the UV and optical, and narrowband \halpha\ 
({\em High Resolution Camera; HRC, and Wide Field Camera; WFC}) to examine
dust, ages in the stellar population, the star cluster population as a whole, and 
estimate the LyC production. 
We use deep ground-based narrowband images in \halpha\ and \hbeta\ in order to
estimate extinction in the gas phase. 
X-ray observations of Haro\,11 have been obtained using the {\em Chandra}
satellite 
\citep{grimes07}
but currently their status is still proprietary. 
We hence also exploit serendipitous off-axis observations from the 
{\em Chandra} and {\em XMM-Newton} X-ray observatories to study the wind properties 
and internal photoelectric absorption.

The article is arranged in the following manner: 
in Section~\ref{sect:data} we describe the observations and data reductions; 
in Section~\ref{sect:res} we present the results;
in Section~\ref{sect:andis} we analyse discuss the results; 
and in Section~\ref{sect:conc} we present our concluding remarks. 
We assume a cosmology of $H_0=72$~km~s$^{-1}$~Mpc$^{-1}$,
$\Omega_\mathrm{M}=0.3$ and $\Omega_\Lambda=0.7$ throughout.
The redshift of Haro\,11 is taken to be 0.020598 from NED, corresponding to a
luminosity distance of 87.1~Mpc.

\section[]{Observations, reductions, and data processing}\label{sect:data}

\subsection{{\em HST} Ultraviolet and optical}

\subsubsection{Observations}

This study makes use of images from all three channels of the {\em ACS} 
onboard {\em HST}: 
the {\em SBC} for the FUV; 
the {\em HRC} for the 2200\AA\ and $U-$band;
and the more sensitive 
{\em WFC} for images in the optical domain. 
The details of the observations and post-reduction processing of the 
{\em HST} UV and optical dataset are described in a companion paper,
\cite{ostlin07}. 
Observations were performed in the bandpasses listed in
Table~\ref{tab:exptime}.
Briefly, {\em F122M} and {\em F140LP} correspond to \lya\ on-line and continuum,
respectively, while {\em FR656N} covers \halpha\ on-line and {\em F550M}
measures line-free continuum bluewards of \halpha. 
The remainder are broadband continuum observations, selected in order 
to avoid the strongest emission lines. 
\begin{table}
\caption{Observations and exposure times}
 \centering
  \begin{tabular}{@{}lllll@{}}
  \hline
  \hline
  Bandpass & &Channel & ExpTime [ s ] & \# Split \\
  \hline
  {\em F122M}   & \lya           & {\em SBC} & 9095 & 5 \\
  {\em F140LP}  &\lya\ cont      & {\em SBC} & 2700 & 5 \\
  {\em F220W}   &$\sim 2200$\AA  & {\em HRC} & 1513 & 3 \\
  {\em F330W}   &$U-$band        & {\em HRC} & 800  & 2 \\
  {\em F435W}   &$B-$band        & {\em WFC} & 680  & 2 \\
  {\em F550M}   &medium $\sim V$ & {\em WFC} & 471  & 2 \\
  {\em FR656N}  &\halpha         & {\em WFC} & 680  & 2 \\
  {\em F814W}   &$I-$band        & {\em HRC} & 100  & 1 \\
  \hline
 \end{tabular}
\label{tab:exptime}
\end{table}

All images are `drizzled' using the {\tt MULTIDRIZZLE} task in 
{\tt IRAF/STSDAS} onto the same pixel sampling scale (0.025\arcsec /pixel) and 
position angle. 
The inverse variance weight maps were saved from the drizzle process since they
provide an estimate of the error in each pixel. 
Remaining cosmic rays, charge transfer tracks, and blemishes are removed from
the CCD observations using the {\tt CREDIT} task.
Remaining band-to-band discrepancies in the astrometric alignment
were rectified using the {\tt GEOMAP} and {\tt GEOTRAN} tasks.
Since our dataset comprises observations between 1200 and 9000\AA\ and utilises
three different {\em ACS} channels, we address the issue of variations in the
point spread function (PSF) of the images. 
PSF models of each band were generated using the {\tt PSF} task in {\tt
DIGIPHOT/DAOPHOT}, with the resulting models being used to convolve all images
to the PSF of {\em F550M} (the broadest emission-line free PSF in our dataset).
PSF models were re-computed for the convolved images and compared to that of 
{\em F550M}; all frames showed PSF full width at half maximum consistent with  
{\em F550M} at below the 5\% level.

\subsubsection{\lya\ continuum subtraction and SED modeling}\label{sect:contsub}

As described in 
\cite{hayes05},
and in more detail in 
Hayes et al. (2007, in preparation), 
continuum subtraction of \lya\ using these filters requires more sophisticated
techniques than most emission lines. 
Because the offline filter is removed from the online filter by $\Delta \lambda
/ \lambda = 0.22$ and the FUV continuum evolves rapidly as a function of
$\lambda$, age, and \ebv, it is imperative to understand the behaviour of the
continuum between {\em F140LP} and {\em F122M}. 
Special care is also required because of the 
broad nature of the {\em F122M} filter which has a rectangular width around 
10\% of the central wavelength, and transmits both the geocoronal \lya\
line and Milky Way \lya\ absorption profile.
In addition, P\,Cygni profiles may result in a reduction of the net emission due 
to the blue-side absorption cancelling some or all of the emission. 

We defined in 
\cite{hayes05}
the {\em Continuum Throughput Normalisation} (\ctn) factor as the factor
that, for a given spectrum, scales the continuum flux sampled in {\em F140LP} to the
flux that would be expected from continuum processes alone in {\em F122M}.
The procedure of estimating \ctn\ in each pixel utilises each of the images 
between {\em F140LP} and {\em F814W}, the filter throughput profiles, and
fitting {\em Starburst99} spectral evolutionary models 
\citep{leitherer99,vazquez05}. 
In 
\cite{hayes05} 
we demonstrated that we could find non-degenerate best-fitting spectra if SED
datapoints sample the UV continuum slope and Balmer/4000\AA\ break. 
That way, for each pixel we fit burst age and \ebv\ using  standard $\chi^2$
minimisation. 
The method has been substantially developed and is described in detail in 
Hayes et al. (2007, in preparation).
We now use continuum-subtracted \halpha\ to first
map the contribution to the overall SED from nebular gas alone.
Thanks to
observations at $V$ and $I$ we are also able to constrain the contribution from 
any stellar population that underlies the current starburst. 
Essentially we measure and subtract the gas spectrum and fit age and mass in
two stellar components, applying the same reddening for all three SEDs. 
For each pixel, we are able to find the age of the stellar populations and \ebv,
treating the nebular gas and two stellar components independently. 
With the best-fitting spectral reconstruction (composite {\em Starburst99} spectra)
at each {\em HST} pixel we 
then map the \ctn\ factor and are able to reliably continuum subtract \lya. 

In Hayes et al. (2007, in preparation) we present extensive simulations,
designed to test the reliability of this methodology for an array of input spectra. 
We determine that the method must account for the presence of an underlying
stellar population and nebular continuum emission.
Both these components may affect the $U-B$ colour and result in bad fits and
poor recovery of \ctn\ in cases where they dominate the optical luminosity. 
Our method of reconstructing the nebular continuum and fitting multiple stellar 
components allows robust recovery of the FUV continuum features and \ctn. 
An independent measure of the metallicity is also found to be a requirement
although the metallicity of Haro\,11 is well known. 
Details of the IMF are found not to have a detrimental impact when 
multiple stellar component fitting  (i.e. a more complex star-formation history) 
is used. 
Overall we find that the software employed here is able to always recover input \lya\ 
equivalent widths to within 30\% for `weak' \lya\ emission (\wlya=10\AA) and to
within 10\% when the \lya\ line is stronger (\wlya=100\AA). 
With regard to the integrated fluxes and visual morphologies, we found in 
\cite{hayes05}
that varying the parameters of the model spectra had a minimal effect on our
results. 
Morphologies were always indistinguishable by eye and integrated \lya\ fluxes
self-consistent to within $\sim 25$\%, even when pushing the parameters 
outside the regimes deemed reasonable for the galaxy in question. 

This procedure also provides numerous other outputs against which we can compare
\lya. 
The full list of output maps is: continuum subtracted \lya\ and \halpha,
\ctn-factor at \lya\ ({\em F122M/F140LP}), continuum flux densities at
900\AA\ (\flyc), 1500\AA\ (\ffuv), and 2200\AA\ (\fnuv), \ebv, the age of the two
stellar components, the mass of the two stellar components, the stellar
equivalent width in absorption of \halpha\ and \hbeta, and $\chi^2$.
These allow us to compare \lya\ fluxes and equivalent widths with various
local stellar ages, \ebv, local star-formation rates, and estimate the
ionising photon production.

Haro\,11 metallicity is known to be around 20\% solar \citep{bergvall02} so for 
this study 
we adopt the evolutionary tracks generated with metallicity $Z=0.004$. Since
we are concerned with OB-dominated, young starbursts we use the tracks of the
Geneva group.
In the absence of a quantitative estimate, the IMF was assumed to follow
that of Salpeter ($\alpha=-2.35$) in the range 0.1~M$_\odot$ to
120~M$_\odot$.
Star-formation history was taken to be that of an instantaneous starburst
(single stellar population) with a mass normalisation of $10^6$~M$_\odot$.

\subsubsection{Binning}\label{sect:reductionbinning}

As described in the introduction, we can expect to see \lya\ features (either 
emission or absorption) around clusters where variations can be expected on small 
scales, and/or large-scale diffuse emission at low surface brightness.
Typically $S/N$ per resolution element is high in the vicinity of strong
continuum sources but significantly lower $(< 1)$ in the diffuse regions. 
Ordinarily, such problems are overcome by smoothing.
While fixed kernel smoothing may improve $S/N$ in diffuse regions it 
smears out spatial details in regions where
$S/N$ is good and smoothing would not be necessary.
Adaptive smoothing may maintain some spatial resolution
but does not necessarily conserve surface brightness.

Adaptive binning provides an alternative to the smoothing approach and for this
study we make use of the Voronoi tessellation code of 
\cite{die06}, a generalisation of the Voronoi binning algorithm of 
\cite{cappel03}. 
Voronoi tessellation overcomes the drawbacks of smoothing by binning
together groups of pixels to conglomerate resolution elements (frequently
referred to as ``spaxels"), recomputing the signal and noise in each new bin. 
Pixels are continuously accreted until a threshold $S/N$ has been met in each
spaxel. 
The advantage of this adaptive binning method is that in high-$S/N$ regions, the
spatial sampling remains high because spaxels are typically small, whereas in
diffuse regions $S/N$ is improved greatly. Diffuse emission regions, by definition, 
show variations over much larger spatial scales.
Since each individual pixel is used exactly once, surface brightness is always
conserved. 

The {\em F140LP} FUV science observation is assigned as the ``reference"
image, and used to generate the binning pattern, using the inverse variance map
output by the {\tt STSDAS/MULTIDRIZZLE} task to compute $S/N$. 
Pixels are accreted into spaxels until $S/N=5$ has been met, with a
maximum size of $40^2$pixels $ = 1\square$\arcsec\ in order to speed up the 
process and prevent spaxels from growing arbitrarily large.

\subsubsection{Super star clusters}

Point-like sources are identified in the deepest of the optical bandpasses, 
{\em F435W} and {\em F550M} using the {\tt DAOFIND} task in {\tt IRAF}. 
To be considered a detection, a cluster must be present in both of these bands. 
Each object was then inspected by eye and any detections that were clearly
spurious were removed from the catalogue. 
For crowded fields PSF-photometry is preferred to standard aperture methods in
order to eliminate cross-contamination from neighbouring clusters.
However, some clusters may be extended enough to be resolved by our 
observations, thus precluding the use of PSF-fitting methods
and limiting us to aperture photometry. 
Aperture photometry was performed using the {\tt PHOT} task in {\tt IRAF} in 
all bands using the {\em F435W+F550M} detected catalogue.
An aperture of 0.10\arcsec\ was used, and sky was sampled in a circular 
annulus of radius between 0.1 and 0.15\arcsec. 
Aperture corrections were then applied in accordance with those given in 
\cite{siri05} for the {\em HRC} and {\em WFC} bandpasses.
While {\em SBC/F140LP} aperture corrections have been computed in the past, 
\citep[e.g.][]{dieball05}, 
they are not available in the published literature and our own aperture
corrections, computed with the {\tt TinyTim} software 
\citep{tinytim}, 
are included here in Table~\ref{tab:apcorr}, Appendix~\ref{sect:apcorr}. 
Since we have no a priori knowledge of the continuum slope, we adopt the {\em
F140LP} aperture correction for $\beta=0$, using 0.1\arcsec\ as in the other
bandpasses. 
Since {\em SBC/F122M} contains the \lya\ line, no point-source photometry 
was performed in this bandpass itself.
Instead, we perform aperture photometry on the continuum subtracted \lya\ and
\halpha\ images in the same 0.1\arcsec\ apertures at the position of each 
cluster with no re-centering applied. 
This gives a direct measure of the \lya\ flux and equivalent width in the
vicinity of each SSC. 

In order to estimate the properties of the SSCs, 
the same SED-fitting software described in Section~\ref{sect:contsub} 
was applied to the aperture extracted fluxes. 
Age, \ebv, photometric mass, etc. were computed, allowing us to
compare these properties with \lya\ in the immediate vicinity of each SSC.

\subsection{ESO -- New Technology Telescope}

Haro\,11 was observed during the nights of 18, 19, and 20 September 2004, using
the {\em New Technology Telescope} at ESO La Silla, as part of an observing run
to obtain \halpha, \hbeta, and [O{\sc iii}] narrowband images for all southern
targets in our {\em HST} \lya\ sample (Atek et al., in preparation). 
On the night beginning 18 Sept seeing was good, not exceeding 1.2\arcsec for the
duration, although thin cirrus prevents a direct calibration of these data. 
On the night of 19 Sept, observational conditions were ideal: photometric 
and with sub-arcsecond seeing throughout. 
The final night was still photometric but seeing deteriorated to $>2$\arcsec\
and images were unusable for science purposes. 
The good-seeing data from the night of 18 was calibrated through secondary
standard stars in the field, using data from the photometric nights of 19 and 20
Sept. 
Narrowband imaging was performed at \halpha, \hbeta, and nearby 
continuum. 
Spectrophotometric standard stars LDS749B, Feige 110, and GD50 were selected
(on the grounds of their high spectral resolution: $1-2$\AA, in order to 
resolve the stellar absorption features around \halpha\ and \hbeta) from the
catalog of 
\cite{oke90}, 
and were 
observed at regular intervals for the duration of each night in all filters.
Imaging observations were performed using both the
{\em ESO Multi-Mode Instrument (EMMI)} 
\citep{dek86} 
and {\em Super Seeing Imager 2 (SuSI2)}
\citep{odor98},
interchangeably.
The instruments were used in $2\times 2$ pixel binning mode to reduce the
readout noise, providing a plate-scale of 
0.1665\arcsec~pix$^{-1}$ 
and 
0.130\arcsec~pix$^{-1}$, 
and a field-of-view (FoV) of 
$9.1 \times 9.9$\arcsec\ and 
$2.2 \times 2.2$\arcsec, 
for {\em EMMI-R} and {\em SuSI2}, respectively. 
Table \ref{tab:ntt} summarises the imaging observations included in this 
article, lists the ESO filters used, and the total exposure times in each band. 

\begin{table}
\caption{NTT observations}
 \centering
  \begin{tabular}{@{}llcc@{}}
  \hline
  \hline
  Observation & Camera & ESO filter \# & ExpTime [ s ] \\
  \hline
   \halpha        & {\em EMMI-R} & 598 &  900 \\
   \halpha\ cont. & {\em EMMI-R} & 597 &  1200 \\
   \hbeta         & {\em SuSI2}  & 549 &  2866 \\
   \hbeta\ cont.  & {\em EMMI-R} & 770 &  1800 \\
  \hline
 \end{tabular}
\label{tab:ntt}
\end{table}

Data were first reduced by the standard routines in {\tt NOAO/IRAF}: 
bias subtraction and flat-field correction using well exposed sky- and
dome-flats. 
Images were aligned using the {\tt GEOMAP/GEOTRAN} tasks and smoothed to the
seeing of the worst seeing image using the {\tt GAUSS} task.
All {\em HST} bandpasses are aligned with the {\em NTT} frames and the
PSF is degraded by convolution with a Gaussian kernel. 
\halpha\ and \hbeta\ were continuum subtracted to obtain the nebular emission
fluxes, accounting for contamination by [\nii] (\halpha\ only) and 
underlying stellar absorption.
[\nii] contamination is estimated using 
[\nii]$\lambda 6583$\AA/\halpha$=0.189$ \citep{bergvall02}.
Stellar absorption is estimated from the best-fitting {\em Starburst99} spectrum
at each pixel by modifying the spectral fitting code described in
Sect.~\ref{sect:contsub}. 
The {\em Starburst99} stellar libraries
\citep{mart05}
were built upon the latest model
atmospheres and include full line-blanketing for all stars and non-LTE effects
for hot stars, thereby providing the best estimate of
the stellar absorption features available.
With the age of the stellar population we measure the equivalent width of
\hbeta\ using the same line and continuum wavelength windows as
the Lick index
\citep{lick}. 
The windows used for \halpha\ are of the same size as those for \hbeta, but
scaled to \halpha\ (i.e. the same windows $\times 6563/4861=1.35$).
\ebv\ is generated from \halpha/\hbeta, assuming a temperature of 10,000~K and
intrinsic line ratio of 2.86, using the extinction law of the SMC following 
\cite{Gordon03,Fitzpatrick_Massa88}.

\subsection{X-Ray: {\em Chandra} and {\em XMM-Newton}}\label{sect:resxray}

Haro\,11 happens to be located 14\arcmin\ away from the Cartwheel
galaxy, of which X-ray observations have been obtained with both the
{\em Chandra} and {\em  XMM-Newton} telescopes
\citep{wolt04}.
Respective total integration times in {\em Chandra} and {\em XMM-Newton}  
were 80 and 70 ksec.   
Regarding {\em XMM-Newton}, Haro\,11 only falls within the field--of--view 
of the {\em MOS2} detector, having unfortunately been missed by both the 
{\em MOS1} and {\em PN} chips.  
Both the {\em Chandra} and {\em XMM-Newton} datasets have been obtained from  
their respective archives.  
Since both telescopes operate with curved focal-plane configuration, the point 
spread function degrades severely with angular distance from the optical axis.  
For the {\em XMM-Newton} detection, the PSF is so distorted that all spatial   
information has been lost, but the object is bright enough that a one-dimensional
spectrum can be extracted from the {\em MOS2} data.   
In the {\em Chandra} observation, Haro\,11 lies on the S1 chip: the object is
clearly detected and we were able to perform a spatial analysis of the source 
(see Sect.~\ref{sect:res_xray}). 

The {\em Chandra} data were processed following the X-Ray Data Centre
pipeline software. 
The level 1 events were reprocessed using {\tt Ciao~3.4} task 
{\tt acis\_process\_events}. 
The whole band (0.2-10 keV) image was smoothed applying the {\tt csmooth} 
task. 
The image was smoothed with a minimal signal--to--noise of 3 using a circular 
Gaussian kernel. 
The source spectrum was extracted from a circular region of radius 30\arcsec.  
The background was obtained from the combination of four circular regions 
located close to the source and avoiding other X-ray sources in the field. 
Source counts were grouped to have at least 20 counts per bin to allow 
modified $\chi^2$ minimisation technique   
\citep{kendall73}  
in the spectral analysis.  
Redistribution matrix and auxiliary response matrix files were generated.

The {\em XMM-Newton} data were processed using the standard  
{\tt Science Analysis System, SAS, v.7.0.0} 
\citep{gabriel04}. 
The most up-to-date files available as of January 2007 were used for the
reduction process. 
The time intervals corresponding to high background events were removed using   
the method followed in   
\cite{piconcelli04}.  
The resulting exposure for the {\em MOS2} data is only of 44 ks. 
No sign of pile-up was found in the {\em MOS2} image, according to the
{\tt epaplot  SAS} task. 
A visual inspection of the 0.2-10~keV image shows a distorted shape due to 
location of the source, close to the edge of the CCD.  
Therefore, no further analysis of the {\em XMM-Newton} image has been performed.  
The source spectrum was obtained, as for the {\em  Chandra} observation, from a 
circular region with a radius of 30\arcsec.   
The background was extracted from a circular region close to the source and avoiding  
other X-ray sources in the  field. 
Source counts were grouped to have at least 20 counts per bin.
{\tt SAS} appropriate tasks were used to generate the distribution
matrix and auxiliary response matrix files.

On-axis observations of Haro\,11 have also been obtained with the {\em Chandra}
telescope
\citep{grimes07}
although their archive status is still proprietary. 
Quantities derived from that dataset are, however, compatible with those presented 
here using off-axis archival observations (see Section~\ref{sect:res_xray}).

\section{Results}\label{sect:res}

\subsection{Lyman-alpha, UV and optical}

\begin{figure*}
\resizebox{0.32\hsize}{!}{\rotatebox{0}{\includegraphics{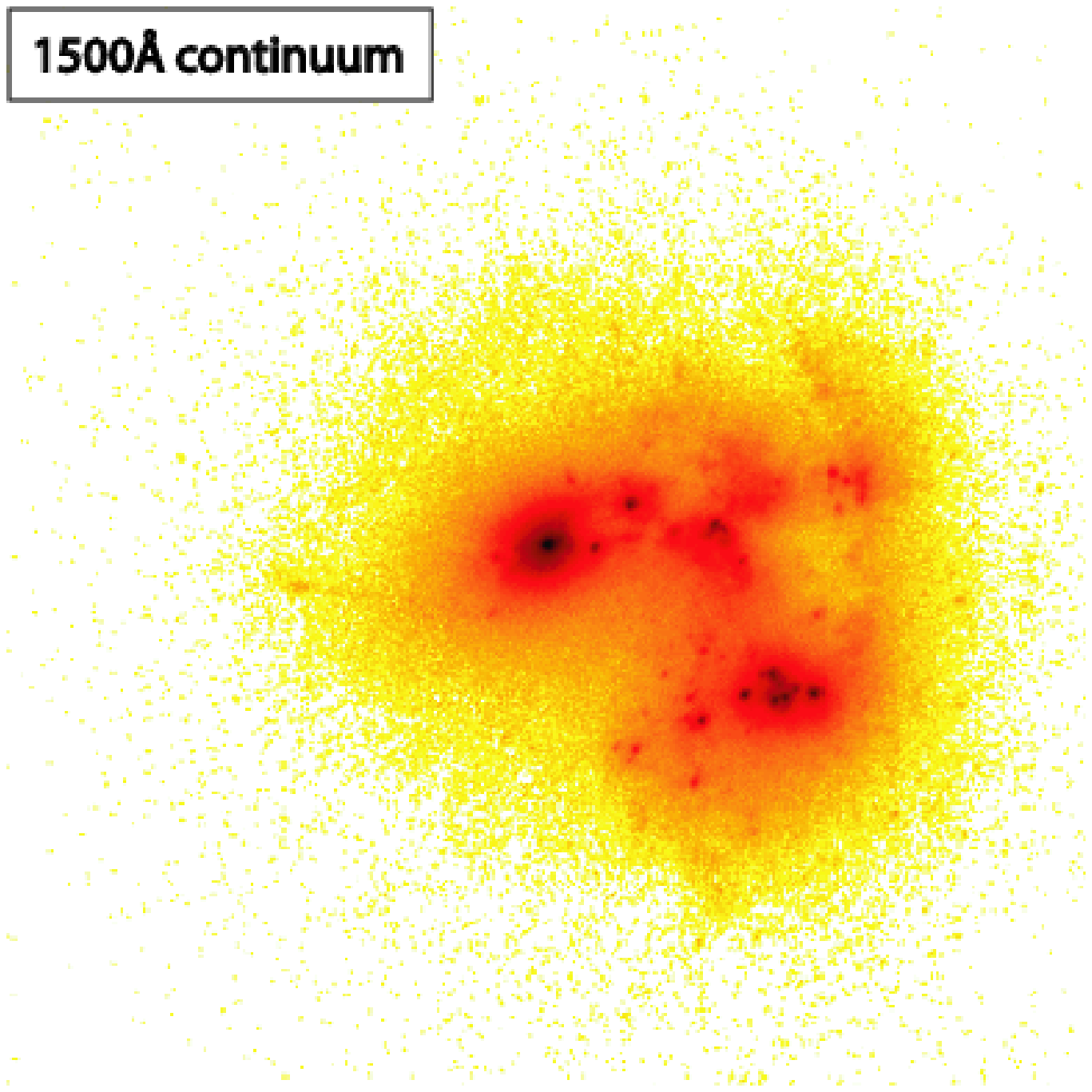}}}
\resizebox{0.32\hsize}{!}{\rotatebox{0}{\includegraphics{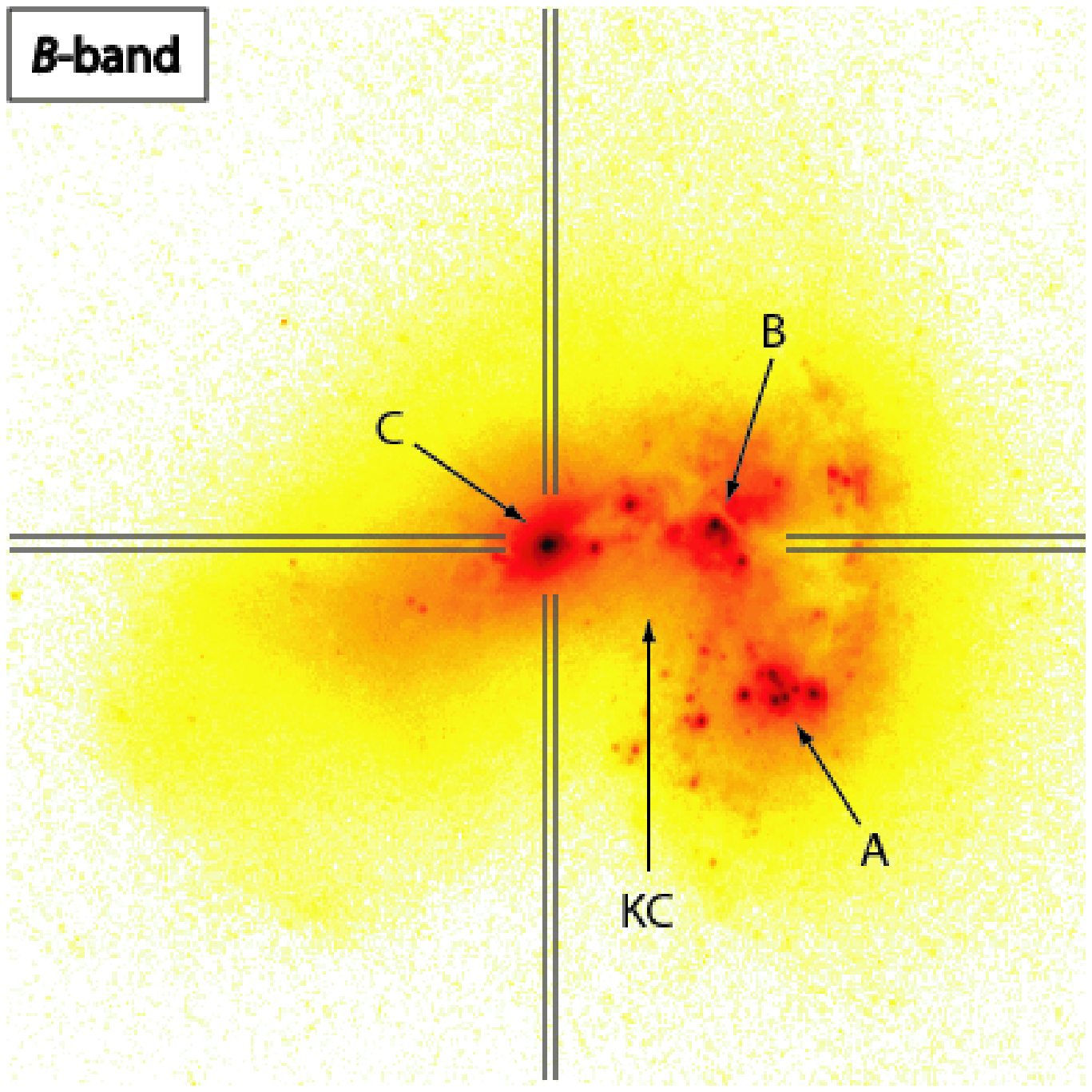}}}
\resizebox{0.32\hsize}{!}{\rotatebox{0}{\includegraphics{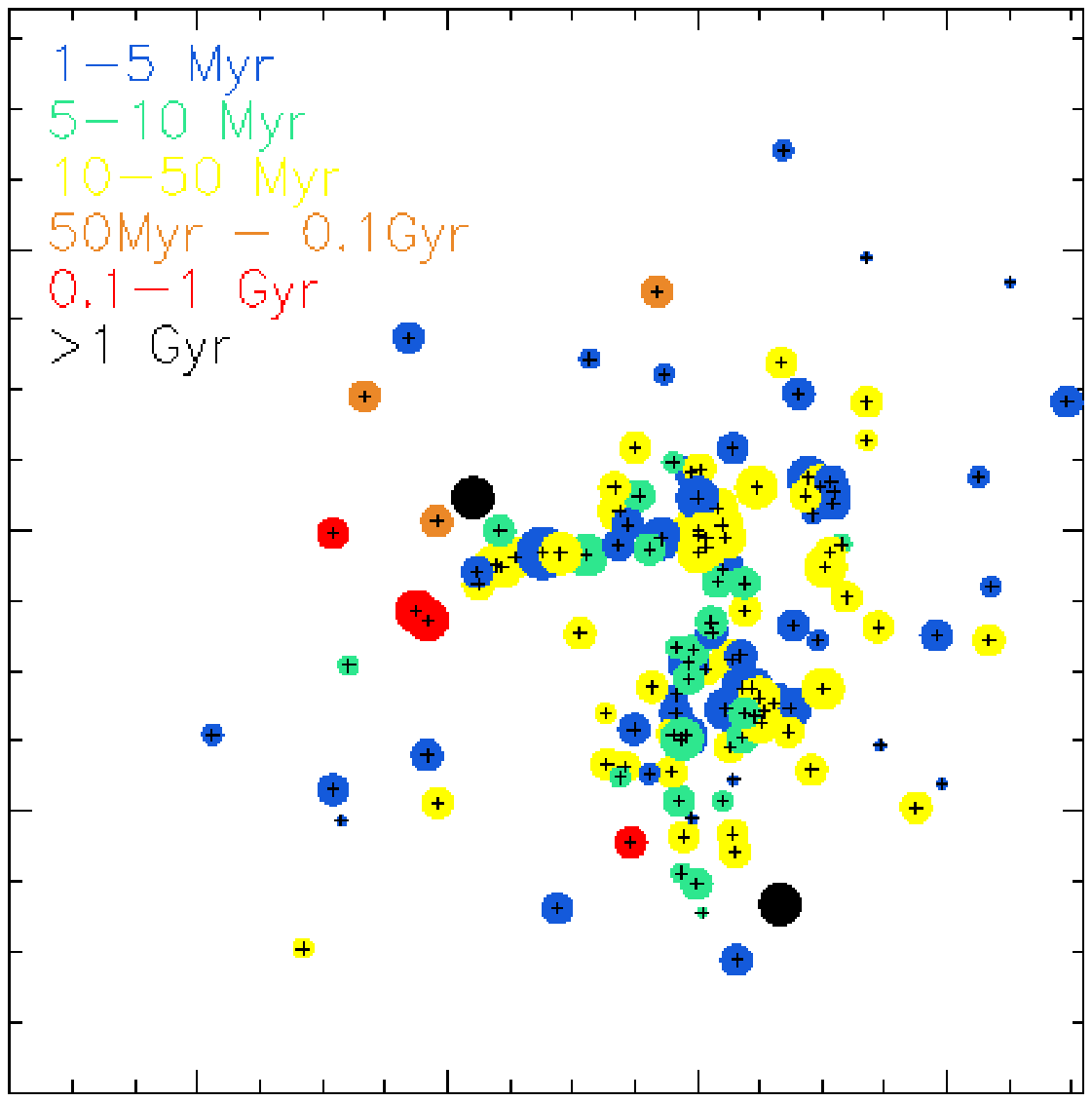}}} \\

\resizebox{0.32\hsize}{!}{\rotatebox{0}{\includegraphics{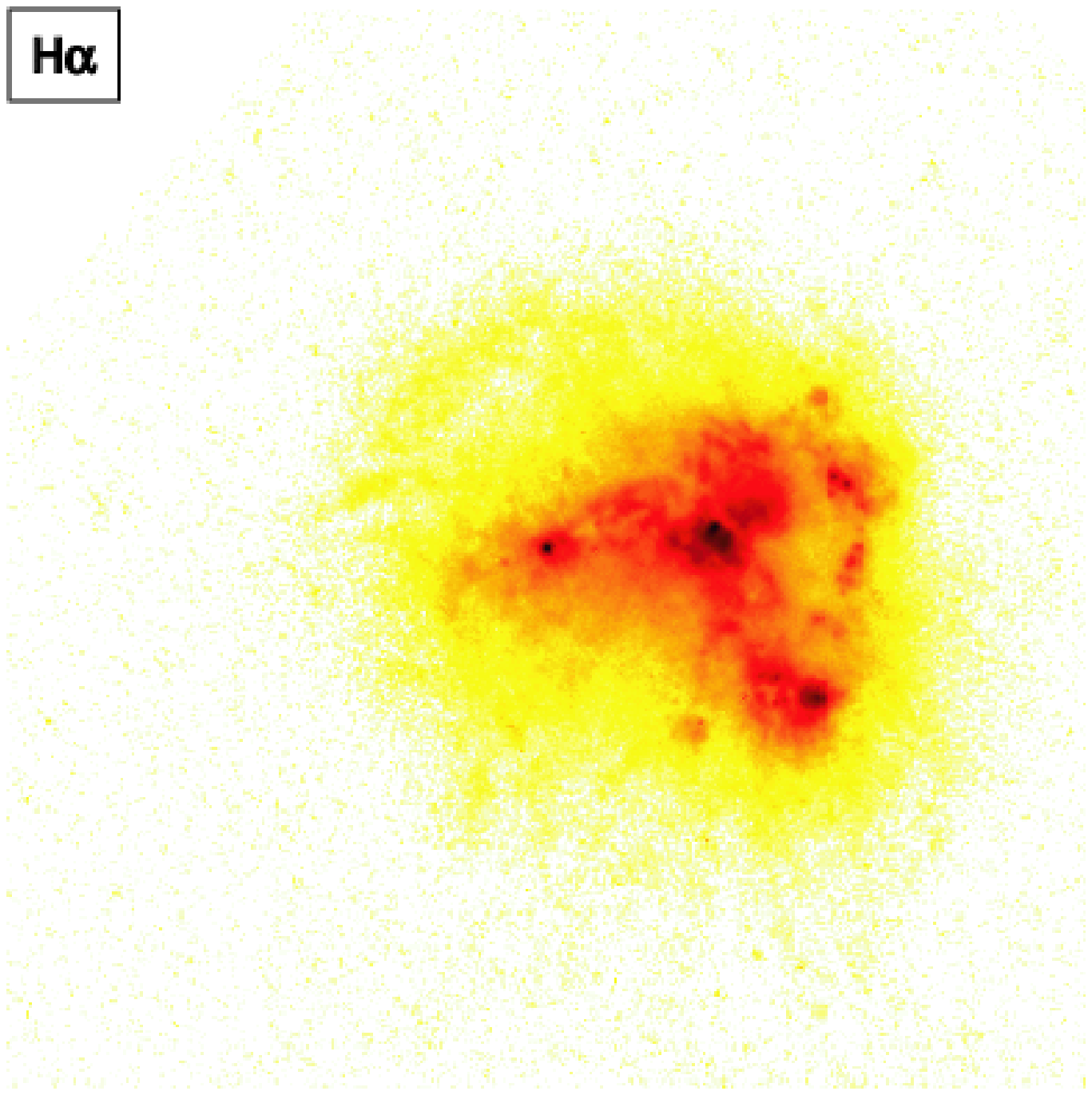}}}
\resizebox{0.32\hsize}{!}{\rotatebox{0}{\includegraphics{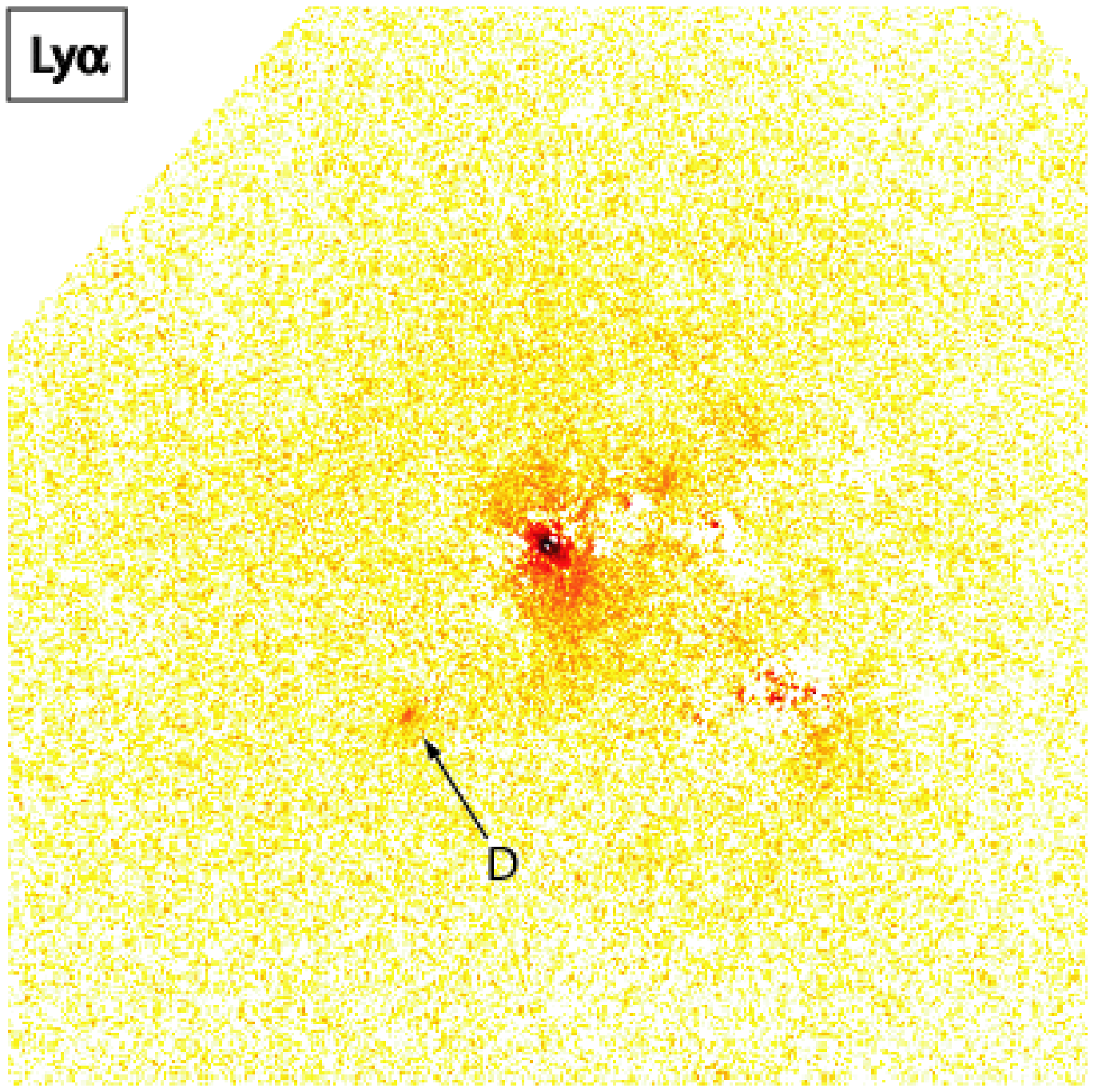}}}
\resizebox{0.32\hsize}{!}{\rotatebox{0}{\includegraphics{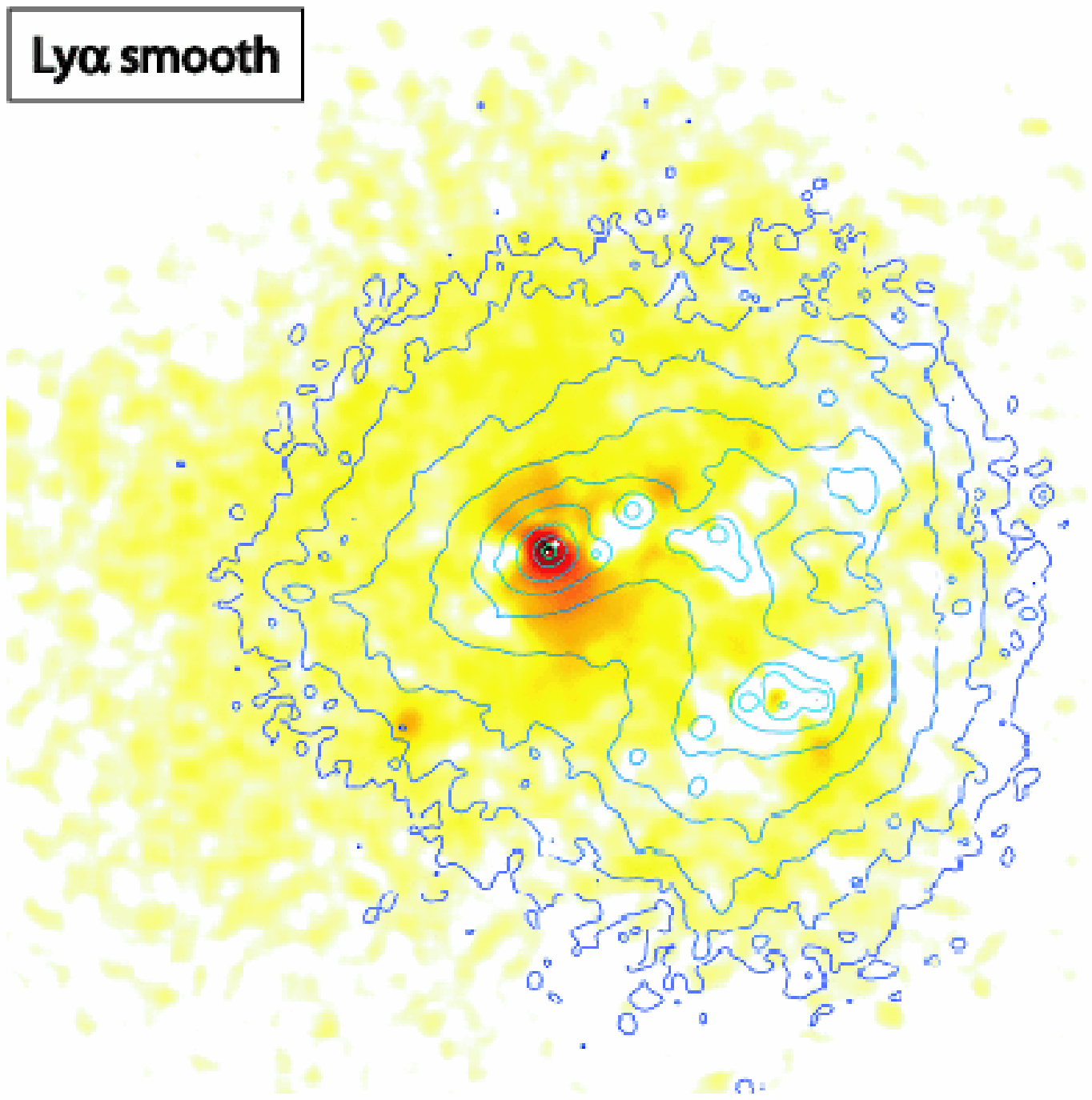}}} \\

\resizebox{0.32\hsize}{!}{\rotatebox{0}{\includegraphics{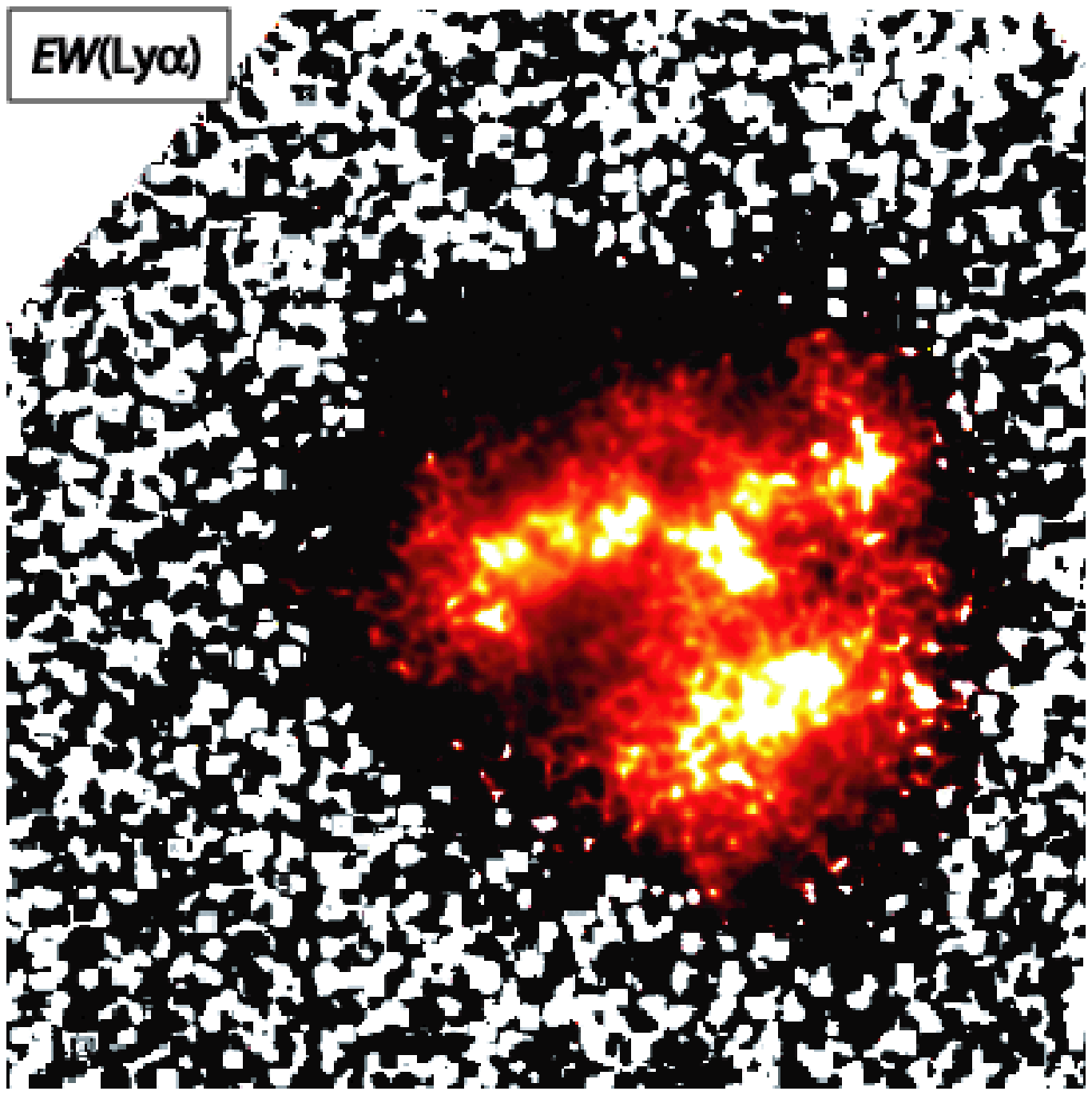}}}
\resizebox{0.32\hsize}{!}{\rotatebox{0}{\includegraphics{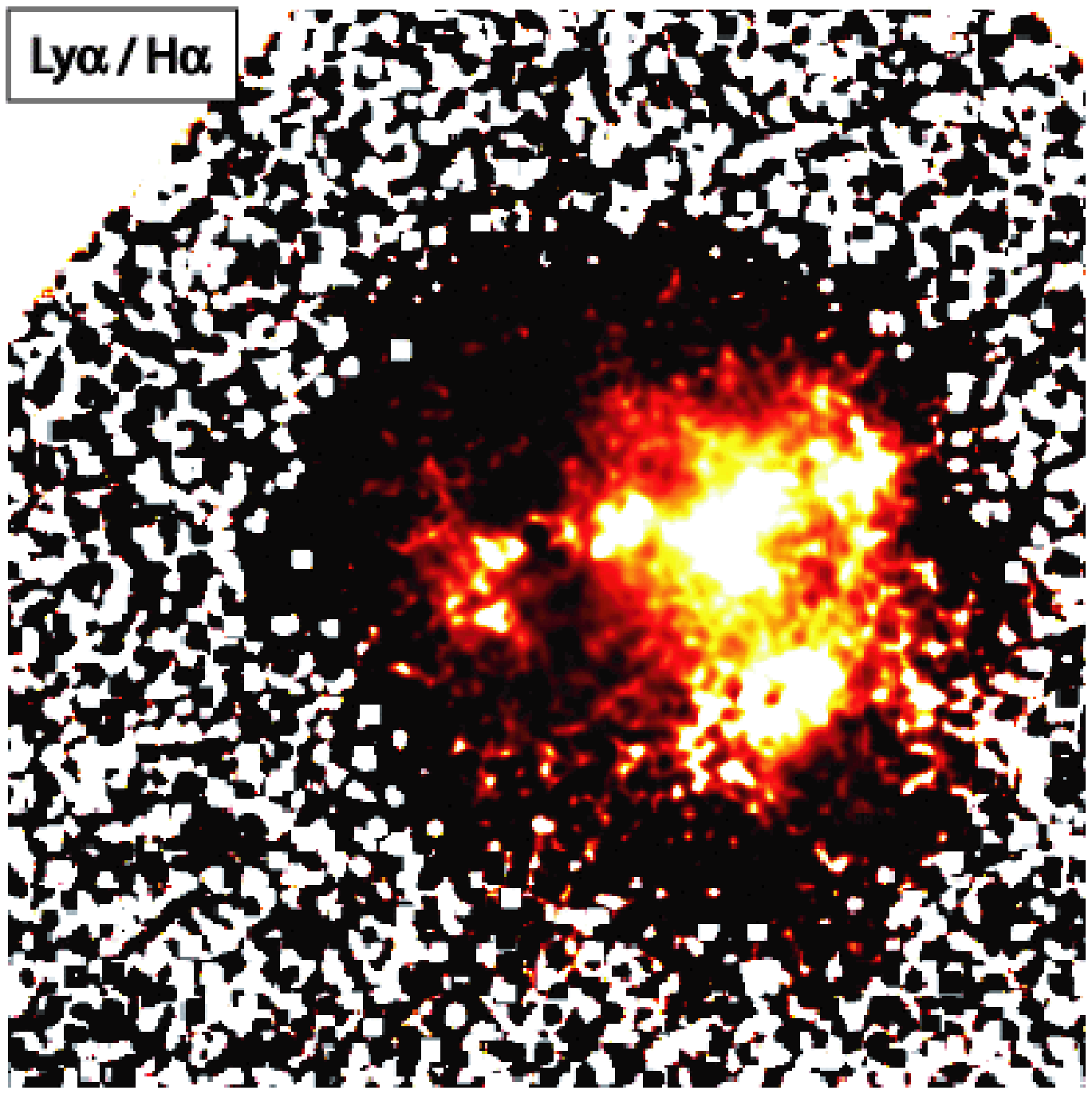}}}
\resizebox{0.32\hsize}{!}{\rotatebox{0}{\includegraphics{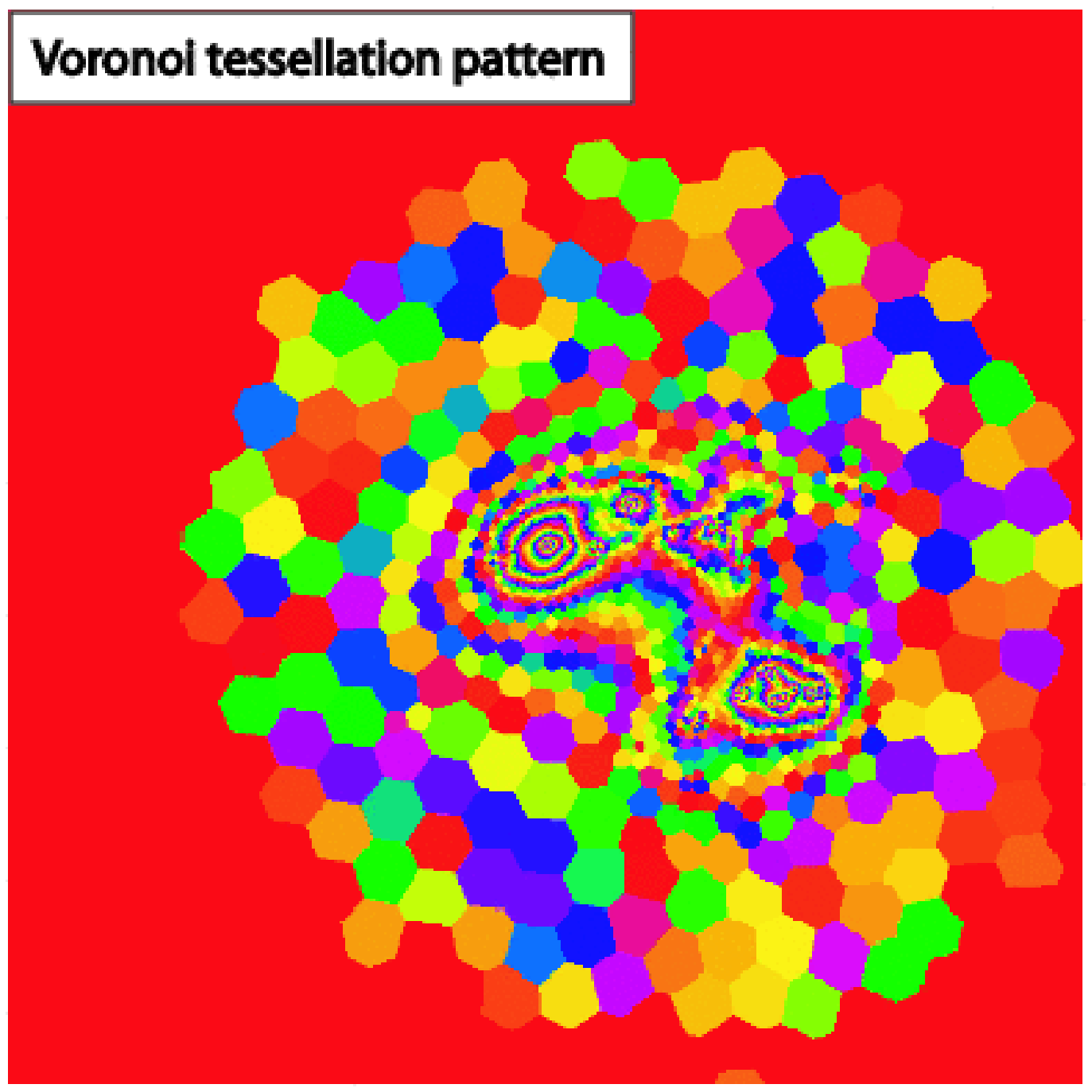}}} \\

\resizebox{0.33\hsize}{!}{\rotatebox{0}{\includegraphics{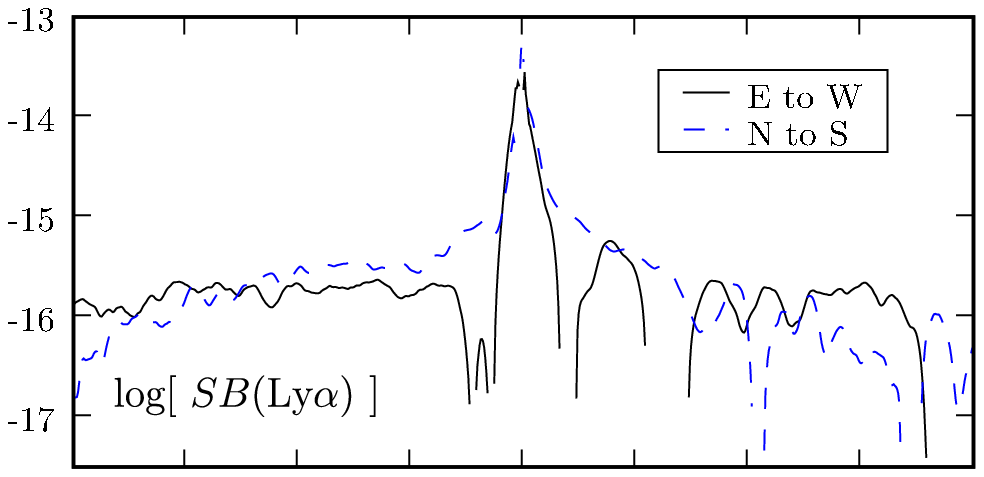}}}
\resizebox{0.33\hsize}{!}{\rotatebox{0}{\includegraphics{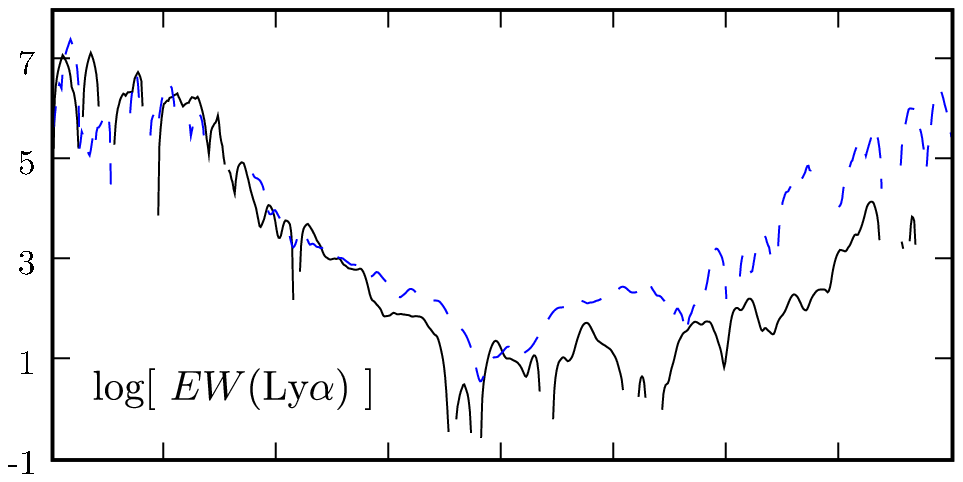}}}
\resizebox{0.33\hsize}{!}{\rotatebox{0}{\includegraphics{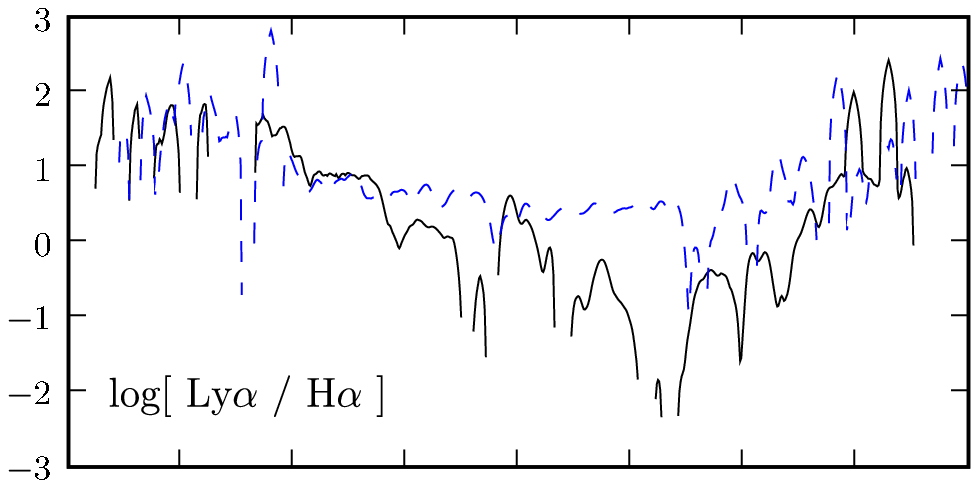}}}

\caption{Haro\,11 as seen by {\em HST}. North is up and east is to the left.
Figure sizes are $20 \times 20$\arcsec, corresponding to a physical size of
8.12~kpc square. 
Images are individually labeled. 
Main star forming condensations and the approximate kinematic centre are labeled in the 
$B-$band image and described in the text. 
The {\em Upper right} image shows the position of the star clusters -- 
size is proportional to $\log(M)$ and colour shows age.
The filtered \lya\ image has adaptively smoothed with
ESO--MIDAS to give $S/N=20$ for an additive noise model and overlayed contours
are from {\em F140LP}. 
Log intensity scale is used throughout with high values showing up in yellow to
black (i.e. black halos in the \wlya\ and \lya/\halpha\ maps represent enhanced
\lya). 
Lower plots show the spatial variation \lya, \wlya, and \lya/\halpha\ in the
rows and columns shown in the $B-$band image.
}
\label{fig:hst}
\end{figure*}

Figure~\ref{fig:hst} shows the results of the processed {\em HST}
observations. 
This panel of figures shows Haro\,11 in FUV continuum and $B-$band, continuum
subtracted \lya\ and \halpha, \wlya, and \lya/\halpha.  
An inverted log-scale is used showing emission in yellow to black. 
Significant star-forming condensations \citep{kunth03} and the approximate
kinematic centre \citep{ostlin99} are marked in the $B-$band image.
Other maps show the position of the SSCs and the Voronoi tessellation pattern.
Plots at the bottom of the figure show the spatial variation of \lya, \wlya, and
\lya/\halpha\ along horizontal and vertical lines through the brightest \lya\
point, the positions of which are marked in the $B-$band image.

We note that \halpha\ emission loosely follows that of the 
FUV continuum (as would be expected from star-formation signatures) 
although there are a few differences. 
Notably, the brightest \halpha\ knot (\B) is not the brightest UV source 
(knot \C). 
This could be an effect of age, dust, the density of the surrounding
gas, or a combination of all three. 
The \lya\ images on the other hand, resemble neither the UV nor 
\halpha. 
Knot \C\ shows strong central \lya\ emission with two other patches of 
more diffuse emission to the north-west. 
Absorption is seen throughout the region of knot \B\ and emission and absorption
can be seen varying on small ($\sim 30$pc scales) around knot \A, although
overall the absorption dominates. 
A compact, isolated emission source to the south-east appears in the continuum
subtracted image and is labeled knot \D.
More generally, however, the \lya\ emission appears in two distinct 
components: a central bright region and a surrounding diffuse, largely 
featureless, halo component. 
In the following analysis we identify these as the {\em central} and 
{\em halo} emission regions as two different physical components. 

Haro\,11 has been observed through the low-resolution {\em Short
Wavelength Prime (SWP)}, $10\times 20$\arcsec\ elliptical aperture of the {\em
IUE} satellite. 
We obtained the {\em IUE} spectrum from the archive and 
have measured the {\em IUE} \lya\ flux to be 
$73.7 \cdot 10^{-14}$~erg~s$^{-1}$~cm$^{-2}$ 
with an equivalent width of 14.8\AA.
Photometry of our \lya\ image in an aperture designed to match that of the {\em
IUE} reveals a flux of 
$75.0\cdot 10^{-14}$~erg~s$^{-1}$~cm$^{-2}$ 
and equivalent width of 14.3\AA. 
This flux is fully consistent with the photometric error on the {\em IUE} measurement 
of around 10\%.

\cite{kunth98} presented \lya\ measurements obtained with the {\em GHRS}. 
This spectrum was obtained 
after a sequence of small angular maneuvers (SAMs), designed to find the peak 
in FUV surface brightness. 
Information about the SAMs is not available and the FUV
morphology is so complex that it is not possible to reconstruct the maneuvers
made by the telescope. 
In a $1.7 \times 1.7$\arcsec\ aperture centred on \C, 
the continuum flux measured in the {\em ACS} frames is a factor of 4 brighter
than that in {\em GHRS}, and apertures centred directly upon \A\ and \B\ both show
\lya\ in absorption. 
We are unfortunately not able to test our {\em ACS} fluxes against this
observation, despite numerous efforts to predict the PEAK-UP behaviour: it's
most likely the telescope centred somewhere between the three knots. 

The total emergent \lya\ line-flux measured in the image 
is 
$79.6\cdot10^{-14}$~erg~s$^{-1}$~cm$^{-2}$,
corresponding to a luminosity of 
$7.22\cdot10^{41}$~erg~s$^{-1}$ 
for our assumed cosmology.

\begin{figure}
\resizebox{0.99\hsize}{!}{\rotatebox{0}{\includegraphics{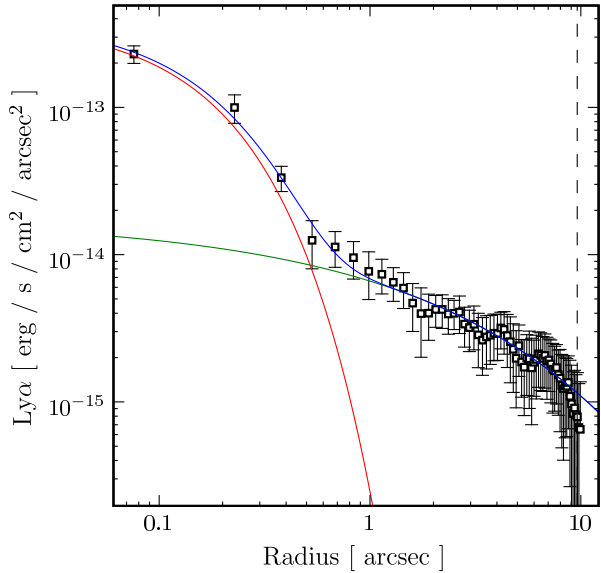}}}
\caption{\lya\ surface brightness profile. 
The vertical dashed line indicates the radius at which the annuli
intersect with the nearest edge of the {\em SBC} detector. 
Green and red lines show the best-fitting S{\'e}rsic profiles to the halo
component, and the central excess, respectively. 
The blue line shows the sum of the surface brightnesses in the two components. }
\label{fig:lya_surfbright}
\end{figure}

To eye the \lya\ image is so azimuthally symmetric that we deem it informative
to present a radial surface brightness profile of continuum-subtracted 
\lya, shown in Figure~\ref{fig:lya_surfbright}. 
This was generated by integrating the flux in concentric annuli of
width 2 pixels, centred on the brightest central pixel of knot \C. 
The central and halo components identified in the \lya\ map are clearly visible 
in the surface brightness profile, with the central component taking over at around 
0.5\arcsec. 
Since they are so clearly independent (visually resembling a bulge+disc
morphology)
we fit S{\'e}rsic functions to the halo and central excess (i.e. total--halo)
components using the formula
\begin{equation}
f_{\lambda, \mathrm{R}} = f_{\lambda, 0} \exp [ -(R/R_0)^{1/n} ] 
\end{equation}
where $f_{\lambda, \mathrm{0}}$ is the central surface brightness, $R_0$ is 
the effective scale length, $n$ the S{\'e}rsic index.
Integration from zero to infinity then gives back the total flux in each
component. 
The results of the profile fitting in the two components can be seen in
Table~\ref{tab:lyasbfit}. 
The regions chosen for fitting in the two components were 0-0.5\arcsec\ 
and 1-7\arcsec for the central region and halo, respectively.

\begin{table}
\caption{\lya\ best-fit surface brightness profile parameters and integrated fluxes}
 \centering
  \begin{tabular}{@{}lccccc@{}}
  \hline
  \hline
  Component & $f_{\mathrm{Ly}\alpha, 0}$  & $R_0$ [ \arcsec\ ] & $n$ &
  $F_{\mathrm{Ly}\alpha}$  & $L_{\mathrm{Ly}\alpha}$\\
  \hline
   Halo           & 2.12 & 1.05  & 2.2 & 76.4 & 6.93 \\
   Central excess & 47.8 & 0.134 & 1.0 & 5.64 & 0.512 \\
   {\bf Total}    &  --  &  --   & --  & 82.0 & 7.44 \\
  \hline
   Measured total &  --  &  --   & --  & 79.6 & 7.22 \\
  \hline
 \end{tabular}
\label{tab:lyasbfit}
\flushleft
Notes: $f_{\mathrm{Ly}\alpha, 0}$, in units 
of $10^{-14}$~erg~s$^{-1}$~cm$^{-2}$~arcsec$^{-2}$.\\
$F_{\mathrm{Ly}\alpha}$ in $10^{-14}$~erg~s$^{-1}$~cm$^{-2}$.\\
$L_{\mathrm{Ly}\alpha}$ in $10^{41}$~erg~s$^{-1}$.\\
$n$ is the S{\'e}rsic index.
Fluxes in the {\em central} and {\em halo} component fluxes are measured by 
integration of the
S{\'e}rsic function from zero to infinity and the {\em Total} flux is the sum of
the two components. The {\em Measured total} is the numerical sum of all the
flux measured in the the annuli.
\end{table}

\begin{figure*}
\resizebox{0.999\hsize}{!}{\rotatebox{0}{\includegraphics{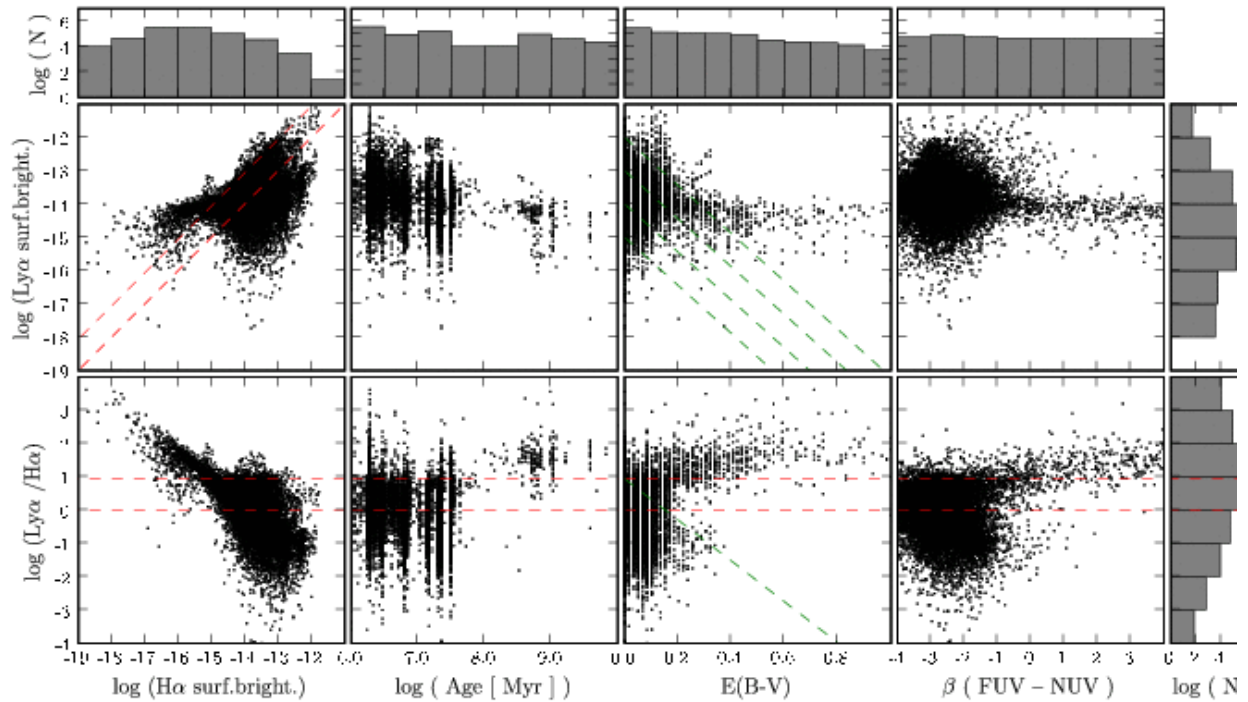}}}
\caption{Resolution element scatter plots of \lya\ surface brightness ({\em
upper}) and the
\lya/\halpha\ line ratio ({\em lower}) against various measured values.
{\em Far left}:  \halpha\ surface brightness.
{\em Centre left}: Age.
{\em Centre right}: Dust reddening \ebv. 
{\em Far right}: FUV continuum slope $\beta$ between 1500 and 2200\AA. 
Red lines show the recombination ratio of \lya/\halpha\ in the case B
approximation and the \lya=\halpha\ line. 
Green lines in the \ebv\ plots show how dust attenuates \lya\ 
for various intrinsic fluxes ({\em upper}) and how it reduces the line ratio 
({\em lower}). 
The resolution elements themselves vary in size and are not representative of
absolute fluxes. 
The histograms above and to the side of the scatter plots show the distribution
of the values, in order to give some impression of how the various parameters
contribute globally.  }
\label{fig:lyavs4}
\end{figure*}

We have also Voronoi binned the individual images.
Since Voronoi binning conserves surface brightness across each spaxel, we can
directly compare \lya\ surface brightness with other fluxes or properties
without a loss of reliability in lower surface brightness regions. 
With binned images in hand, we present the \lya\ surface brightness and 
\lya/\halpha\ as a function 
of various other parameters in Figure~\ref{fig:lyavs4}. 
Each point represents the surface brightness in each resolution element,
although the elements themselves vary from single pixels to 1600-pixel
(1 $\square$\arcsec) spaxels.
Also, since fluxes are presented on log-space, only resolution elements
that are positive in \lya\ are shown, eliminating regions of absorption.
It is not possible to estimate global or integrated values
from these scatter-plots; histograms of the distributions are presented above 
and
to the side of each scatter plot, in order to better give a feeling for this.
The left-most two plots show how the \lya\ emission and line ratio correlate 
with \halpha\ and the age of the young stellar population. 
The \lya\ {\em vs.} \halpha\ plot is again suggestive of two distinct pixel 
components.
One at lower \halpha\ surface brightness ($\lesssim
10^{15}$erg~s$^{-1}$~cm$^{-2}$~arcsec$^{-2}$) in which \lya\ appears largely
independent of \halpha. 
This feature extends over approximately three orders of magnitude in \halpha\ and the
\lya/\halpha\ plot shows the emission is always above the value that would
be expected from recombination theory.  
Above the quoted \halpha\ surface brightness, the distribution changes and an
upper limit appears to be set by the recombination value. 
The line ratio plot clearly shows that \lya\ photons are emitted from regions where 
the \halpha\ flux would suggest that little nebular gas is present.
A similar effect is also seen in the second plot where \lya\ is compared with the age of the
stellar population and there is evidence for substantial amount of pixels
showing \lya\ emission with ages in excess of 100~Myr; much too old to produce nebular
emission. 
The right-most two plots show how \lya\ correlates with the \ebv\ as determined
from fitting the stellar UV continuum (\ebvs), and continuum slope itself (\bet).
Since \bet\ in part reflects \ebv, these two plots 
correlate somewhat in shape, although \bet\ is also dependent upon age. 
The discrete values of \ebvs\ from the fitting grid are apparent.
Again the distribution signifies a non-continuous distribution of pixels: in both cases
\lya\ can be seen in emission from regions with \ebv\ extending up to 1, and
$\beta$ showing a very red slope. 
In these regions of the plots, emission is again found almost exclusively above
the recombination value. 
At lower \ebv\ ($\lesssim 0.3$) and $\beta$ ($\lesssim -1$) the pixel
distribution changes and broadens significantly in both \lya\ and 
\lya/\halpha.
The \lya\ surface brightness tends to increase in these regions 
although the bulk of the pixels show line ratios significantly below the 
recombination value.
While brighter overall, these emission regions show strong attenuation of 
\lya. 

\begin{figure}
\resizebox{0.98\hsize}{!}{\rotatebox{0}{\includegraphics{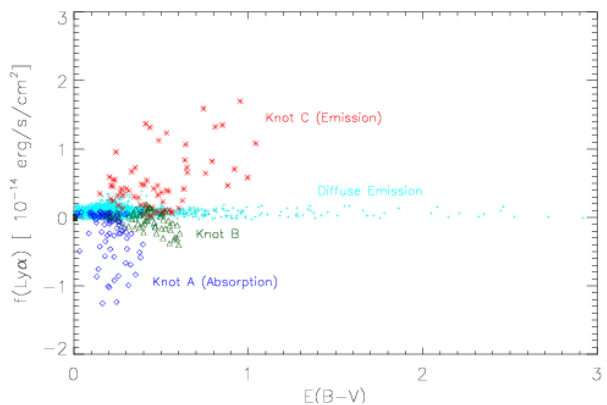}}}
\caption{ \flya\ {\em vs.} interstellar \ebv\ (\ebvg) as determined by the Balmer
decrement. Regions of knots \A, \B, \C, and the diffuse emission component are
represented in different colours and marked in the figure. 
}
\label{fig:lya_ebv_ground}
\end{figure}

Figure~\ref{fig:lya_ebv_ground} shows how \flya\ correlates with gas-phase
interstellar extinction, \ebvg,
as determined from the Balmer decrement and is partly similar to the 
right-most plots in Figure~\ref{fig:lyavs4}. 
This provides a much less biased estimate of dust extinction compared to the
approximation obtained by spectral modeling of the stellar continuum.
Table~\ref{tab:ebvabc} shows the values of \ebvg\ obtained by measuring \halpha\
and \hbeta\ in 1\arcsec\ circular apertures centred on knots \A, \B, and \C,
together with the values obtained spectroscopically by 
\cite{vader93}. 
With the spatial sampling in {\em HST} images degraded by two orders of magnitude to 
match the ground-based seeing, we cannot resolve the fine details of the \lya\ 
emission and the surface brightness range is significantly reduced. 
\flya\ is here presented in linear space and now shows regions of absorption, most
notably around knot \A. 
This distribution shows a structure partially resembling that of the \lya\ {\em vs.} 
\ebvs\ plot: a component of diffuse, low surface brightness \lya\ emission is seen
extending beyond \ebvg$\sim 1$, with a brighter \lya\ emission region (knot \C)
where \ebvg\ shows a large spread, with a mean value of 0.48.
\begin{table}
\caption{\ebvg\ as measured in the {\em NTT} \halpha\ and \hbeta\ frames,
averaged in circular apertures centred on the 3 main star forming
condensations. 
Values obtained by Vader et al. (1993) are shown for comparison. }
 \centering
 \begin{tabular}{@{}llll@{}}
 \hline \hline 
 & \A\ & \B\ & \C\ \\
 \hline  
 {\em NTT} \halpha/\hbeta & 0.20 & 0.42 & 0.48 \\
 \cite{vader93}           & 0.16 & 0.41 & 0.39 \\
 \hline
\end{tabular}
\label{tab:ebvabc}
\end{table}

\begin{figure}
\resizebox{0.98\hsize}{!}{\rotatebox{0}{\includegraphics{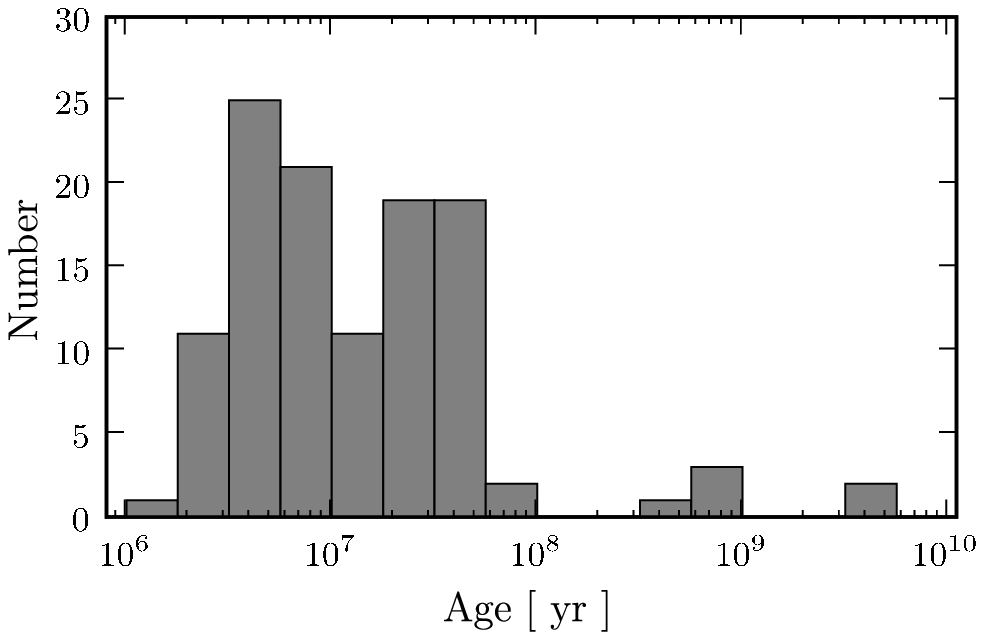}}}
\resizebox{0.98\hsize}{!}{\rotatebox{0}{\includegraphics{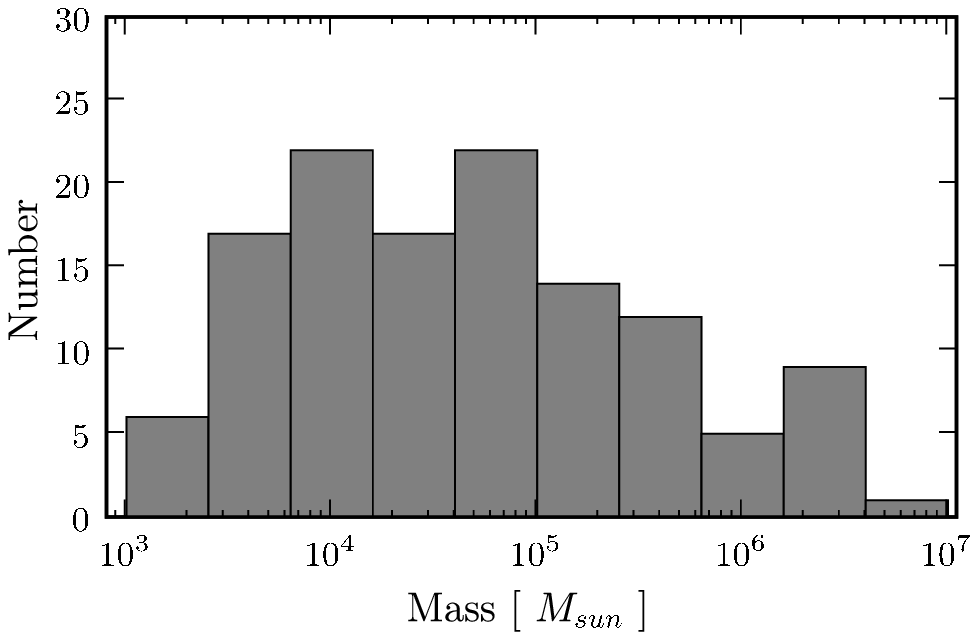}}}
\caption{{\em Upper}: Age distribution of the star clusters; 
{\em Lower}: Mass distribution. Knot \C\ is dominated by a single SSC with a mass
of $2.2 \times 10^6$~$M_\odot$, with an age of around 2~Myr.}
\label{fig:ssc_ages}
\end{figure}

In total we find 140 star-clusters.
The spatial distribution, together with the ages and masses of these clusters
can be seen in the upper-right panel of Figure~\ref{fig:hst}. 
The age and mass distribution of the clusters can be seen in 
Figure~\ref{fig:ssc_ages}. 
The decrease in the mass distribution at $\lesssim 10^4M_\odot$ is the result of
incompleteness in the less luminous clusters.
While there is much of interest in these data, most of it is outside the scope of
this article and will be presented in a forthcoming publication (Adamo et al., 2007 in
prep); here we simply deal with the relation between a few basic properties and
\lya.

\subsection{X-ray: {\em XMM} and Chandra}\label{sect:res_xray}

\begin{figure}
\resizebox{0.98\hsize}{!}{\rotatebox{0}{\includegraphics{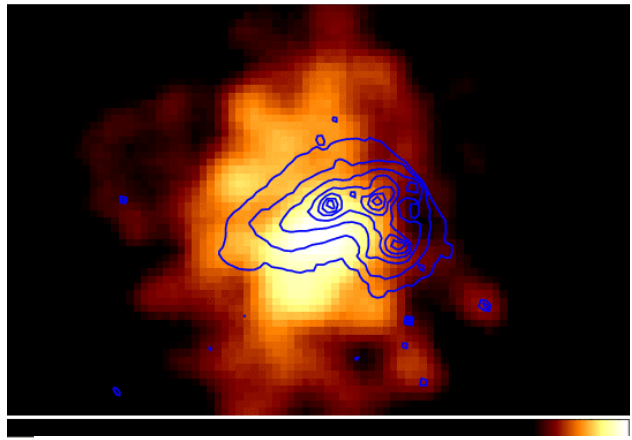}}}
\resizebox{0.95\hsize}{!}{\rotatebox{0}{\includegraphics{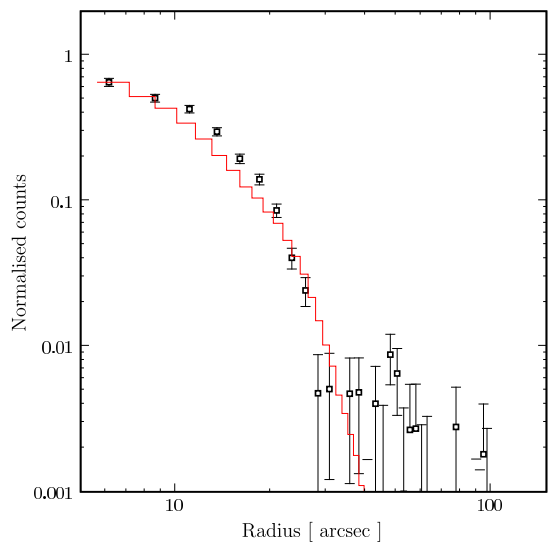}}}
\caption{{\em Upper}: Haro\,11 as seen by {\em Chandra} over the complete
0.3-8keV spectral range. 
Image is $40$\arcsec$\times 25$\arcsec.
Overlaid are the {\em FUV} contours as seen by {\em HST/F140LP}.   
{\em Lower}: {\em Chandra} X-ray source surface brightness
profile over the 0.2-2~keV range together with the corresponding PSF at 1.5 keV.
Datapoints consistently exceed the PSF between $\sim 10$ and 20\arcsec, showing
the source to be resolved and spatially extended.  }
\label{fig:chandra}
\end{figure}

In the upper panel of Figure~\ref{fig:chandra} we show the {\em Chandra} 
image of the full energy range (0.3--8~keV) adaptively smoothed 
(see Sect.~\ref{sect:resxray}). 
Continuum contours from the {\em ACS/F140LP} observation have been overlaid 
to show the location of the different knots. 
The lower panel in Figure~\ref{fig:chandra} shows the X-ray surface
brightness profile, in the 0.2--2~keV range compared with the
corresponding profile of the off-axis PSF of {\em ACIS-S1} at 1.5 keV. 
The plot shows hints of extended emission visible in the 20--40~px region,
i.e. $\sim$10-20~\arcsec\ (1 pix = 0.492\arcsec).
The {\em Chandra} image shows a central unresolved core region and a weak, diffuse 
hot gas component, emitting over an area $\le 30$\arcsec\ in diameter.
The diffuse component appears more extended in the north-–south direction than in 
east–-west. 
However, no firm conclusion can be reached as the image is largely distorted due to 
the $\sim 14$\arcmin\ offset from the {\em Chandra} nominal on-axis location. 
In fact, the {\em Chandra} on-axis images of Haro\,11 reported by 
\cite{grimes07}, 
shows different emission knots following the optical/UV emission structure.
The brightest peak of the 
\cite{grimes07} 
image coincides with knot \B, while the maximum emission on our image is located 
closer to knot \C\ and more than 5\arcsec\ from knot \B.
A plausible explanation for the observed mismatch could be variability, 
although the spectral analysis, fluxes, and luminosities of 
both observations are in good agreement (see below).
Therefore, the more likely explanation could be the intrinsic {\em Chandra}
astrometry uncertainty, although we note that this is only of the order of
2\arcsec\ for off-axis observations.

\begin{figure}
\resizebox{0.99\hsize}{!}{\rotatebox{-90}{\includegraphics{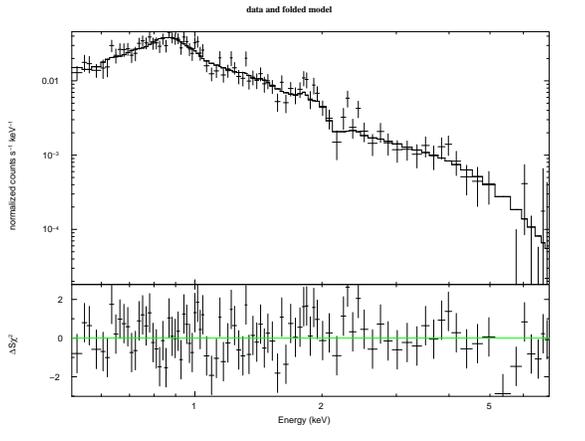}}}
\caption{{\em Chandra} spectral extraction in the 0.2-10~keV range, the best
  fit  model, and  the residuals.  Model  parameters can  be found  in
  Table~\ref{tab:xraypars}.}
\label{fig:chandraspec}
\end{figure}

Spectral analysis was performed on both {\em Chandra} and
{\em XMM-Newton} data, in order to derive the spectral shape of the source,
the X-ray luminosities and, local photoelectric absorption column
density of H{\sc  i}.   
The observed spectrum was corrected for Galactic absorption, 
\nhi $= 2.39\cdot10^{20}$cm$^{-2}$
\citep{kalberla05,bajaja05}. 
The Haro\,11 spectral emission is complex: both power law and thermal 
single--component emission models provide unsatisfactory fits to the data,  
with $\chi^2_\nu$ values equal to 2.2 and 3.5, respectively. 
The best-fit model consists of a combination of a power-law and {\em mekal} 
model 
\citep{mewe86}.
Statistically indistinguishable results were obtained applying a double  
thermal {\em mekal} model.  
However, the temperature of the hottest thermal model is $9^{+4}_{-2}$~keV.  
This high value is difficult to explain in a standard physical scenario. 
We have also derived an upper limit for the absorption equivalent Hydrogen column
of $6\times10^{20}$~cm$^{-2}$.   
Figure~\ref{fig:chandraspec} shows the X-ray spectrum together with the best fit model 
and the corresponding residuals.  
The values of the parameters and the goodness of the fit are listed in  
Table~\ref{tab:xraypars}. 
The corresponding total X-ray fluxes and unabsorbed luminosities are shown in
Table~\ref{tab:xrayflux}.   
Even though the effective area of {\em  XMM-Newton} is larger than that 
of the {\em Chandra}, the {\em MOS2} spectrum of Haro\,11 resulted in 
lower signal--to--noise. 
This is  mainly due to the short real exposure time (44~ks) after
the high background filtering, combined with the fact that the source
is only visible in {\em MOS2}, for which the effective area is 3 times 
lower than that of {\em PN}. However, the source MOS spectra was analysed,
confirming the results obtained with {\em  Chandra}.    
A single--component model (power-law or thermal emission) failed to
reproduced the data. 
The best fit  model consists also of a power-law with $\Gamma\sim2$ and a  
{\em mekal} component with $kT\sim0.4$.  
All the parameters derived from the {\em XMM-Newton} spectrum are 
compatible within the error with those from the {\em Chandra} analysis. 
We have measured a somewhat lower value of the high energy flux,  
of the order of 65$\pm20$\%.
The quantities derived from our spectral analysis are compatible with those
derived from on-axis {\em Chandra} observations presented in 
\cite{grimes07}. 

\begin{table}
\caption{Best-fitting parameters of X-ray spectral model. }
 \centering
 \begin{tabular}{@{}lll@{}}
 \hline \hline Parameter & Value & Unit \\
 \hline  Intrinsic $n_{HI}$ & $<6\times10^{20}$ & cm$^{-2}$ \\
 Power-law photon index & $1.72\pm0.18$
	 \\ {\em Mekal} kT  & $0.69^{+0.08}_{-0.04}$ & keV \\
 \hline
\end{tabular}
 \flushleft
  $\chi^2$: 95.30 using 88 PHA bins.\\
   Reduced $\chi^2$: 1.08 for 88 degrees of freedom. \\
  \label{tab:xraypars}
\end{table}

\begin{table}
\caption{Observed X-ray fluxes and unabsorbed luminosities.}
 \centering
  \begin{tabular}{@{}lll@{}}
   \hline
   \hline
    Band [ keV ] & Flux  & Luminosity \\
   \hline
   0.5 to 2  & $  9.7 ^{+0.4} _{-1.3} $ & $ 1.00 ^{+0.04} _{-0.13} $\\
   2 to 10   & $ 13.3 ^{+0.9} _{-1.5} $ & $ 1.20 ^{+0.08} _{-0.14} $\\
   \hline
   Total     & $ 23.0 ^{+1.4} _{-2.0} $ & $ 2.26 ^{+0.14} _{-0.21} $\\
   \hline
 \end{tabular}
 \flushleft
  Flux in $10^{-14}$~erg~s$^{-1}$~cm$^{-2}$ \\
  Luminosity in $10^{41}$~erg~s$^{-1}$
\label{tab:xrayflux}
\end{table}

\section{Analysis and discussion}\label{sect:andis}

\subsection{Lyman-alpha morphology}

Figure~\ref{fig:hst} shows some of the imaging results. 
Visual inspection clearly reveals that \lya\ morphology is not representative of
either \halpha\ or the FUV continuum.
The overall \lya\ morphology exhibits a central bright emission source (centred
on knot \C), with a
diffuse, largely symmetric halo-like component with lower surface
brightness. 
The emission around \C\ is more extended in the north--south
directions compared to east--west, and appears as two fan-like structures of 
size $\sim 1$\arcsec.
This structure is not seen in \ebv\ maps and may be due to a filamentary 
\hi\ structure.
Typically \wlya\ and \lya/\halpha\ are low in the central regions, where
continuum and \halpha\ are strong, but increase dramatically in the outer
regions. 
Some absorption spots are seen in the regions around knots \A\ and \B. 
The most likely interpretation of this overall morphology is that photons are 
escaping directly from the central regions, either through winds or ionised holes
in the ISM, while the halo component arises from resonance scattering of
\lya.

Knot \D, the small blob of \lya\ emission found in the analysis of 
\citep{kunth03} 
is clearly visible towards the south--west. 
Knot \D\ is not visible in any of the continuum bands or the \halpha\ on-line
observation and it appears to be a real source of \lya\ photons devoid of continuum. 

The radial surface brightness profile presented in Figure~\ref{fig:lya_surfbright} 
shows the flux in the halo component decaying approximately exponentially with 
radius, but peaking sharply in the centre.
The fan-like structures described above are contained within the central surface
brightness peak.
Fitting S{\'e}rsic functions to these two independent components and integrating
under the best-fitting formulae allows us to measure the relative contribution
of the components of central and halo emission, and shows that while surface
brightnesses are significantly lower, $\sim90$\%
of the photons are emitted via the photon diffusion mechanism.
This is comparable to the value of $\sim70$\% in the halo component
of ESO\,338-IG04 presented in 
\cite{hayes05}, 
although for that study the estimate was made by masking the images instead of
profile fitting. 
The two galaxies are comparable in FUV luminosity, both low-metallicity, and
both exhibit complex dynamical and morphological structures, and the \lya\
output of both are clearly dominated by the halo mechanism.

\subsection{Lyman-alpha production, escape and regulation}

In the \halpha\ image we measure a total line flux of 
$333 \times 10^{-14}$~erg~s$^{-1}$~cm$^{-2}$, identical to that 
given in 
\cite{schmitt06}, which is not corrected for Galactic extinction. 
Galactic extinction correction 
\citep{ccm89}
gives a intrinsic \halpha\ flux of 
$342 \times 10^{-14}$~erg~s$^{-1}$~cm$^{-2}$, consistent with 
\cite{ostlin99}. 
The total \lya\ production estimated from the case B recombination
gives a value of 
$2975 \times 10^{-14}$~erg~s$^{-1}$~cm$^{-2}$.
Adopting the \lya\ flux we obtained by integrating the (Galactic-extinction
corrected) surface brightness profile we obtain a 
{\em \lya\ escape fraction of just 2.7\%}. 
This calculation neglects the fact that \lya\ can be enhanced relative to \halpha\ 
by shock excitation. 
Were there a significant contribution to \lya\ as a result of a shocked ISM,
this escape fraction will be further reduced. 

Plots of the \lya\ surface brightness in each binned element, compared 
with various other quantities, are seen in
Figure~\ref{fig:lyavs4}.
Most importantly, these plots show how the \lya\ surface brightness spatially
compares with \halpha\ and the dust distribution. 
The left-most plot shows \lya\ {\em vs.} \halpha; the red lines indicate
\lya=\halpha\ line and \lya/\halpha=8.7 (case B recombination). 
Comparing \lya\ to \halpha\ also identifies two independent components of \lya\ 
emission.
Firstly, at low \lya\ and \halpha\ surface brightness ($\lesssim 10^{-14}$ and
$10^{-15}$~erg~s$^{-1}$~cm$^{-2}$~arcsec$^{-2}$ for \lya\ and \halpha,
respectively), the \lya\ to \halpha\ ratio seems to always exceed that predicted
by recombination (see \lya/\halpha\ spatial distribution in
Figure~\ref{fig:hst}). 
Again, shock excitation could enhance the production of \lya\ relative to
\halpha, and it's plausible that some of the \lya\ is produced in a shocked ISM.
Indeed, this is an oft-proposed mechanism for \lya\ production in
high-$z$ \lya-blobs (LABs), resulting in large-scale \lya\ emission with
little or no associated continuum.
However, without observations in other emission lines the exact mechanisms at
play in LABs are very difficult to test, although tests may be performed for
targets at low-$z$ observed with {\em HST}.
Optical spectroscopy of Haro\,11 performed by 
\cite{heisler_vader94}
and 
\cite{kewley01} 
consistently find spectra consistent with pure photoionisation nebulae.
On the other hand, these spectra were extracted in large apertures centred on the
knots, avoiding diffuse \lya\ emission regions, and the answer is not fully 
conclusive. 
It is possible that future high resolution spectroscopy (e.g. with a repaired
{\em STIS}) may be able to test this directly.
Given the resonant nature of \lya\, a more likely interpretation of this
\lya--\halpha\ decoupling is that it arises from 
resonant scattering of \lya\ photons.
The \lya/\halpha\ histogram in Figure~\ref{fig:lyavs4} confirms
that over 50\% of the pixels are in the component with super-case B line ratios.
The fact that these points are here most likely indicates that a significant number 
of \lya\ photons must have travelled far from their point of origin, undergoing 
numerous resonance scatterings. 

At higher surface brightnesses 
($\gtrsim 10^{-15}$~erg~s$^{-1}$~cm$^{-2}$~arcsec$^{-2}$ for \halpha), the
distribution clearly changes. 
Here the recombination value seems to set an upper limit on the \lya/\halpha\
ratio and \lya\ is never stronger than expected. 
The top edge of this distribution must result from \lya\ photons escaping the
galaxy directly either due to ionised holes in the ISM cleared by the UV radiation
field or through a rapidly outflowing neutral medium.
The distribution fans out in this component and a large spread in \lya\ surface
brightness is seen which results from either resonance scattering
selectively removing \lya\ photons from the line-of-sight (from where they may
be re-scattered and emitted in the halo component) or destruction by dust. 
The central emission region also shows absorption structures to the east
and west of knot \C\ (see middle-right panel of Figure~\ref{fig:hst}), where
this effect may be strongest.

The resonant decoupling of \lya\ photons is also revealed in
Figures~\ref{fig:lyavs4} and \ref{fig:lya_ebv_ground}, where \lya\ is compared
with \ebv\ from fitting to the stellar continuum (\ebvs), \halpha/\hbeta\ (\ebvg), 
and \bet.
If \lya\ were a non-resonant line a declining relationship would
be expected between \lya\ and \ebv. 
The bulk of the \lya\ output is in the halo component 
(\lya\ surface brightness 
 $\lesssim 10^{-14}$~erg~s$^{-1}$~cm$^{-2}$~arcsec$^{-2}$;
Figures~\ref{fig:lya_surfbright} and \ref{fig:lyavs4}).
Below this level, there is essentially no trend of \lya\ with \ebvs. 
Most importantly, when \lya\ is seen in emission from regions with
\ebvs$\gtrsim0.2$ it is almost entirely with line ratios that exceed the
recombination value. 
This uncorrelated faint emission component is also seen in the \ebvg\ plot 
where the distribution of diffuse \flya\ pixels (shown in cyan) is constant 
over the whole range of \ebvg.
Despite the low dust content of knot \A\ (mean \ebvg$=0.2$), \lya\ is seen only in
absorption in this region. 
In contrast, the brightest \lya\ region (knot \C) shows \lya\ only in
emission with measured values of \ebvg\ as high as 1, although
the regions \A, \B\ \& \C\ have sizes comparable to the seeing
disc in the ground-based images and therefore the scatter at each
knot may be an artefact (the mean value of \ebvg\ in knot \C\ is 0.48).
\lya\ reaches its brightest flux levels where \ebvs\ is small which is
symptomatic of the contrasting escape mechanisms,
but it must be noted that this only accounts for around 10\% of the overall
\lya\ output. 
That we see such strong absorption from knot \A\ but emission from knot \C\ seems 
to 
indicate that the static H{\sc i} column density in the line-of-sight to knot
\A\ 
must be significantly greater than to knot \C. 
Knot \A\ does contain several massive clusters, the ages of which are consistent
with the high mechanical luminosity 
\citep{leitherer99}
that would be required to drive out the neutral ISM, although it seems that
either the H{\sc i} covering is too great, or the clusters are too young to have
accelerated the surrounding medium. 
Unfortunately we have no direct information on the H{\sc i} column density or
the mechanical energy return that would be required to drive a superwind.

That we see \lya\ from resolved regions with \ebv$>0.5$ is a striking result.
For a simple dust slab this would correspond to the loss of more than 
99.9\% of \lya\ photons, before scattering is considered.
Since in the halo regions we typically see \lya/\halpha\ above the
recombination value we can infer that radiative transfer effects are at play in
the propagation of \lya\ photons, and 
scattering events have caused these photons to travel significantly 
increased path-lengths.
Again, knowledge of the geometry and kinematics of the neutral and ionised 
media would be required to infer the number of scatterings, mean free 
paths, and therefore the actual path-lengths involved. 
We can, however, identify two possible extremes of the photon transport
mechanism. 
Firstly, they may have diffused through largely static \hi\
gas with transport dominated by multiple rare scatterings into the wings of the 
distribution function. 
In this case, the neutral medium must be pristine, dust-free, and have never
mixed with star-forming regions. 
Secondly, and much more likely, much of the propagation may have been through
paths in the low-density ionised ISM with scattering/reflection events occurring
at the surfaces of dense neutral clouds. 
In the multiphase ISM models of 
\cite{neufeld91} 
and 
\cite{hansen06}, 
such a geometry shields \lya\ from the dust that is embedded in the \hi, with
reflections preserving \lya\ by preferential propagation in the ionised ISM.
Transport effects in these models suggest that equivalent widths may be significantly 
boosted, even globally. 
The global \lya\ equivalent width and line-ratio are certainly not boosted, and
what we are seeing may be preferential \lya\ destruction in some regions and
relatively unhindered transport in others. 
Where the bulk of the \lya\ destruction is taking place, we cannot say. 
Either way, the transport effects result in the superposition of \lya\ photons
on top of dusty and/or old stellar populations.

A two dimensional kinematic study of Haro\,11 has been performed in the \halpha\
 line by 
\citet{ostlin99,ostlin01} 
using the {\em CIGALE} Fabry-Perot (FP) interferometer on ESO's 3.6m telescope. 
These data demonstrated the kinematic centre to lie approximately
midway between knots \A\ and \C\ as labeled in Figure~\ref{fig:hst}, with the major
axis lying in the NW-SE direction with the NE side toward knot \B\ receding.
In the central 2\arcsec (1kpc) iso-velocity contours are shown to be densely 
packed and show a steep rise in rotational velocity in the direction of knot \B,
indicating a non-relaxed system or possible second kinematic component. 
While we expect even the diffuse \lya\ emission to be in part regulated by the 
kinematic structure of the ISM, we are not able to see any correlation between
\lya\ and velocity shifts or shear in the ISM.
The resolution of the {\em CIGALE} data is not sufficient for us to compare
small-scale \lya\ emission/absorption features with the kinematic structure of
the ISM, although on larger ($\sim$kpc) scales no correlation is apparent. 
Moreover, turbulence or disruptions in the ionised medium does not necessarily 
imply the neutral ISM has been accelerated.

In summary we see clear evidence that, while \lya\ photons must be produced in the 
same regions as \halpha, we do not observe them to be emitted from their
production sites, and the bulk of the \lya\ photons travel significantly
enhanced path lengths. 
We see no relationship that would indicate \lya\ correlates spatially with the
dust distribution. 
Unfortunately we do not have information about the H{\sc i} distribution or its
kinematic structure.

\subsection{Lyman-alpha and the SSC population}

\begin{figure}
\resizebox{0.9\hsize}{!}{\rotatebox{0}{\includegraphics{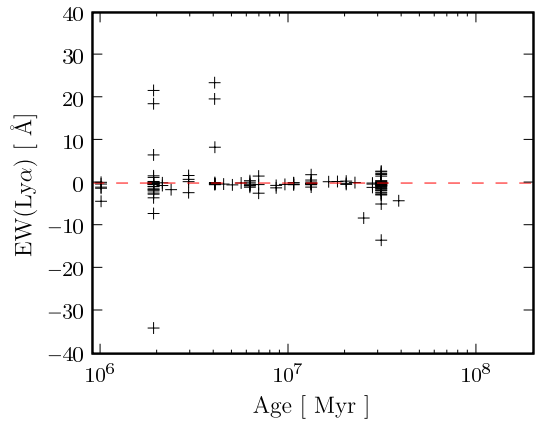}}}
\caption{Distribution of \lya\ equivalent width in the immediate vicinity of the
each super star cluster compared with the cluster age.}
\label{fig:ssc_lya}
\end{figure}

The age distribution  of the super star cluster
population are shown in Figure~\ref{fig:ssc_ages}. 
Figure~\ref{fig:ssc_lya} shows the equivalent width of \lya\ in
the immediate 0.1\arcsec\ of the super star clusters, 
compared with the respective age of each cluster. 
\lya\ equivalent widths are measured from the continuum-subtracted
and continuum-only image in fixed apertures. 

\cite{tt99} and 
\cite{mashesse03}
have presented evolutionary studies of \lya\ emission from starbursts which
qualitatively predict the evolution of \wlya\ with age. 
The \lya\ sequence is more complex than that of \halpha\
due to the
resonant trapping of \lya\ photons in H{\sc i} resulting in a need for mechanical 
feedback and/or a higher ionisation fraction to create a medium optically thin to
\lya. 
The model is such that at very young ages ($\lesssim 1.5$~Myr) the  \lya\ feature
is expected in absorption due to dense, static \hi\ medium.
After some time, the cluster is able to ionise this medium and drive out an
expanding shell resulting in a pure emission feature between 2.5 and 5~Myr.
As the expanding shell recombines, the \lya\ emission feature develops a
P\,Cygni-like profile which, with continuing  recombination removes successively 
more of the emission until, at around 8~Myr, the \lya\ feature is
again in absorption. 
With such a large number of SSCs (effectively mini galaxies) and \lya\ maps, 
this evolutionary sequence becomes something we could possibly examine within our 
dataset. 
However, there are a number of factors which could mask such a relationship, 
particularly in a dataset consisting of imaging observations only. 
Firstly, when the feature is P\,Cygni some or all of the emitted flux is
cancelled by the absorption profile, resulting in termination of the observed \lya\
emission phase at earlier times than predicted.
Secondly, geometry also enters into the model and, 
due to the ionisation structure being cone-like, the
orientation must be favourable in order to observe the \lya\ emission phase. 
In our dataset, this will result in a decrease of \wlya\ or complete absence of
\lya\ from some \lya\ emitting clusters. 
Thirdly, due to the diffuse emission, \lya\ can be seen from
essentially any region of the galaxy. 
However, due to the high continuum levels 
around SSCs, strong contamination is not expected and will also be accounted
for by background subtraction. 
Finally, a general spread can be expected due to variations in the initial
cluster mass, \hi\ distribution, mechanical feedback requirements, and
internal absorption that is not local to the cluster itself. 
What we could possibly expect is negative \wlya\ at the earliest times, becoming
positive and peaking between 2.5 and 4~Myr, and then declining again to the
background levels at around 6 to 8~Myr. 
Spread is expected toward lower (and negative) \wlya\ due to orientation effects
and the \hi\ covering.
Moreover, the noisy nature of the continuum-subtracted \lya\ map further
complicates very local equivalent width measurements. 
We consider only the sub-population of clusters with age below 40~Myr. 
Figure~\ref{fig:ssc_lya} does show a distribution of peak \wlya\ 
{\em vs.} age that is quantitatively consistent with the evolutionary models.
For the lowest ages \lya\ is seen only in absorption or around the zero-level.
For ages between 2 and 4~Myr some SSCs show local \lya\ emission that could
possibly represent the emission phase, although there is some spread. 
At late times ($\gtrsim 8$~Myr) \lya\ is only seen in absorption or very close to
zero.

\subsection{X-ray information and star formation rates}

The {\em Chandra} image (Figure~\ref{fig:chandra}) shows that  most of the
X-ray photons have been emitted in a region surrounding knot \C.
The radial profile of X-ray emission is largely featureless.
Comparing the radial profile with the {\em Chandra} off-axis PSF
(Figure~\ref{fig:chandra}) yields some hints of extended emission up to around
10--20\arcsec.
The X-ray emission arises from the same region around knot \C\ where
\lya\ is brightest, suggesting that \lya\ emission from this region might be the
result of a perturbed and outflowing ISM, accelerated
by the release of mechanical energy from the SSCs located at knot \C.
However, it is worth noting that the analysis of an on-axis {\em Chandra}
observation of Haro\,11 
\citep{grimes07}
indicates that the emission is
resolved in several knots, the brightest one coinciding with knot \B. 

We have measured an upper limit for the equivalent Hydrogen absorption column of
$6 \times10^{20}$~cm$^{-2}$. We want to stress that consistent fits could also
be found with lower columns: the F-test gives only a probability around 90\% for
the fits using this value of the column density, so that the present data do not
allow us to constrain the amount of neutral Hydrogen in front of  knot \C\ . In any
case, column densities around $2 \times10^{20}$~cm$^{-2}$ were obtained  by
\cite{kunth98} from Voigt profile fitting of the \lya\ absorption profile in the
{\em GHRS} spectrum ($n_\mathrm{HI} = 2.5 \cdot10^{20}$~cm$^{-2}$), and
\cite{verhamme06} note that Voigt profile fitting may systematically
underestimate \nhi. It is interesting to note that Haro\,11 showed the largest
\hi\ column density in the 4 galaxies of \cite{kunth98}   with \lya\ emission
($\log(n_\mathrm{HI}) = 19.7 - 20.4$ cm$^{-2}$). The derived  temperature  of 
the  hot diffuse  component is  $\sim 8\cdot10^{6}$K,  a   typical  value  also 
found for other local dwarf-starbursts like He\,2-10 and NGC\,4449
\citep{grimes05}.

\begin{table}
\caption{Integrated star-formation rates in Haro\,11. The VLA data have
  been  taken from Condon et al. (1998). } 
  \centering
   \begin{tabular}{@{}lcll@{}}  
   \hline 
   \hline  
   Tracer &  SFR &  Source & Calibration  reference\\   
   \hline  
   1.4~GHz    &  5.6            & {\em VLA}   & \cite{condon92} \\  
   FIR        &  18             & {\em IRAS}    & \cite{rosagonzalez02} \\
   \halpha\   &  24             & {\em HST}    & \cite{kennicutt98}  \\ 
   1527\AA\   &  19             & {\em HST}    & \cite{rosagonzalez02}  \\ 
   0.5-2  keV &  22.0$^{+0.9}_{-3}$ & {\em Chandra} & \cite{ranalli03} \\  
   2-10 keV   &  26$^{+2}_{-3}$ & {\em Chandra} & \cite{ranalli03} \\

  \hline \end{tabular}
  \flushleft
 SFR [ M$_\odot$~yr$^{-1}$ ] 
 \label{tab:sfr}
\end{table}

In Table~\ref{tab:sfr} we present star-formation rates (SFR) derived from
X-rays, other observations in this study, and previous observations from the
literature.  Within the errors, star formation rates derived from FUV continuum,
Far Infrared ({\em IRAS} FIR, corrected to the 1--1000 $\micron$ range 
according to
\cite{helou88}),
\halpha\ and X-ray luminosities are consistent
with a SFR $\sim20-25$~M$_\odot$~yr$^{-1}$. Such a SFR is rather high and could
explain the presence of high-velocity winds and disrupted interstellar medium,
driving the \lya\ escape and pushing hot and neutral gas to a large radial
extent. 
Compared to the sample of \lya\ emitters at $z\sim3$ from 
\cite{gronwall07}, 
Haro\,11 appears at the top end of the SFR distribution they derive from the 
FUV continuum (within the range 1--40, with an average value around 
5~M$_\odot$~yr$^{-1}$). 
This implies that Haro\,11 is a very actively star-forming galaxy in the local 
Universe, comparable to the objects being analysed at high redshifts and 
the results obtained from its analysis may therefore be directly applicable 
to more primeval systems.

The fact that the SFRs derived from the X-ray emission are pretty consistent
with the values derived from FUV continuum, \halpha\ and  FIR emission implies that
the X-ray emission should be driven by the powerful star formation processes
taking place in Haro\,11, with no hint on the presence of a low luminosity active
nucleus at the core of knot \C.  
Indeed, the $L_{0.5-2 keV}/L_{FIR}$ ratio of $2.6\cdot10^{-4}$ is very close 
to the average value found by \cite{ranalli03} 
for a sample of star-forming galaxies at redshifts in the range 0.2--1.3. 
A preliminary analysis based on the models of \cite{oti07} 
shows that this ratio is consistent with a powerful star
formation process dominated by SSCs with age below 5~Myr, with a standard
efficiency of around 10\% in the transformation of released mechanical energy
into thermal X-ray emission.

On the other hand, the radio luminosity  yields a significantly lower value of
the star formation rate when compared to the other tracers. We believe this is
due to the youth of the star formation episodes in Haro\,11. The calibration used
assumes a steady star-formation process with radio contributions from both
thermal and non-thermal components (synchrotron emission from supernova
remnants). Figure~\ref{fig:ssc_ages} shows that most of the SSCs are younger 
than around 3.5~Myr, so that the number of supernovae expected from the slightly 
older SSCs is significantly smaller. The low radio emission could therefore be 
explained by this deficit in the non-thermal component.

The relative hardness of the X-ray continuum (photon index $\sim1.7$),
indicates that the hard band, contributing $\sim50$\% of the X-ray output,
might be dominated by high-mass X-ray binaries (HMXB). Since HMXBs become
active after 3-4~Myr of evolution, we believe that these sources could be
already present in the SSCs with age between 4 and 10~Myr.  
Indeed, an SSC of mass $\sim 1.1\times10^6$~$M_\odot$ with an age of 8.2~Myr
identified near knot \C\  could be already hosting a significant number of HMXB.
The
observational properties are therefore consistent with the distribution of
ages shown in Figure~\ref{fig:ssc_ages}: old enough to have produced the number of HMXBs
seen in the hard X-ray bands, but still young enough not to have generated so many
SNe that the radio luminosity would be dominated by synchrotron emission.

\subsection{Lyman continuum production and emission}

{\em FUSE} observations 
\citep{bergvall06} 
have shown Haro\,11 to be an emitter of LyC radiation, with an observed flux
density at 900\AA\ (\flyc$_{,\mathrm{e}}$) in the 30\arcsec$\times$30\arcsec\ 
aperture  of $1.1 \times 10^{-14}$~erg~s$^{-1}$~cm$^{-2}$~\AA$^{-1}$.
Subsequent re-analysis of the data
\citep{grimes07}
place \flyc$_{,\mathrm{e}}$ at a lower value of 
$0.23 \times 10^{-14}$~erg~s$^{-1}$~cm$^{-2}$~\AA$^{-1}$ although the answer is
not yet conclusive. 
\cite{bergvall06} 
estimated the intrinsic LyC production by summing the measured LyC leakage 
and an estimate of the ionising flux computed from \halpha\ and
recombination theory. 
This translated to an escape fraction of between 4 and 10\% assuming a Salpeter
IMF, although the higher value is reduced to $\sim 2$\% in the analysis of 
\cite{grimes07}. 
However, this method of estimating \flyc$_{,0}$ is difficult to
calibrate for a given wavelength since \hi\ is ionised by all photons with
$\lambda < 912$\AA, not just those at 900\AA. 

As described in Section~\ref{sect:contsub}, our software also outputs maps of the
intrinsic \flyc, based upon the normalisation of best-fitting population
synthesis models. 
This relies upon stellar continuum and allows us to predict directly the value 
of \flyc$_{,0}$ at each pixel, thereby
giving a different, and possibly more robust handle on the ionising photon
budget than that estimates based upon \halpha\ alone: 
\halpha\ estimates provide and upper limit on the stellar ionisation budget
since shock excitation of \halpha\ is neglected, while stellar continuum fitting
cannot account for ionising photons that arise from non-thermal processes. 
However, in light of the tight correlation between between X-ray determined SFR
and SFRs determined by other methods, we do not expect a significant non-thermal 
contribution.

\begin{figure}
\resizebox{0.95\hsize}{!}{\rotatebox{0}{\includegraphics{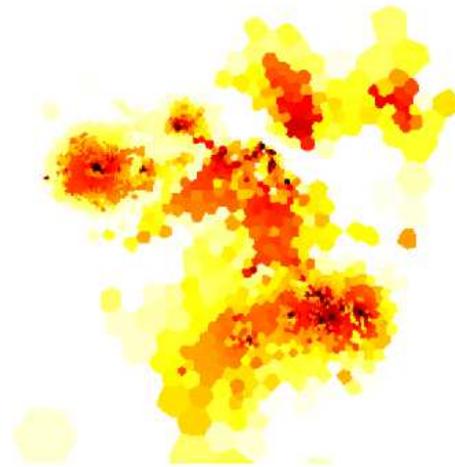}}}
\caption{Production map of the flux density at 900\AA. North is up and east to
the left. 
Image dimensions are $12 \times 12$\arcsec, and concentrates only on the central
star-forming regions where LyC photons can be expected to be produced.  }
\label{lyc}
\end{figure}

Figure~\ref{lyc} shows a map of the ionising photon production, zoomed in on the
central star-forming regions (i.e. the sites of LyC production). 
As expected, the LyC morphology itself resembles that of \halpha, with knots \A,
\B, 
and \C\ showing up as strong production sites. 
Knot \B\ is the brightest of the three in LyC, consistent with indications from
the \halpha\ that it is the most intensely ionising source. 
An apparent dust lane is seen north--west of \B\ which is seen in the FUV
continuum and optical bands, and may represent the failure of the implementation
of `dust screen' extinction laws to recover all the intrinsic 900\AA\ flux when 
dust is distributed along the line--of--sight.
Since the method of computing \flyc\ relies on the normalisation of 
model spectra, \flyc\ can never be negative and
pixel-to-pixel noise can introduce a spurious positive contribution to the
integrated flux. 
Here we deal only with the Voronoi-binned images in order to better control 
the background noise.

Summing the 900\AA\ flux we obtain a total value of
$f_{900,0}=12.3\times10^{-14}$~erg~s$^{-1}$~cm$^{-2}$~\AA$^{-1}$. 
This translates into a 900\AA\ escape fraction, $f_{\mathrm{esc}}$, of 9\%. 
This is towards the upper end of the range presented in \cite{bergvall06} but,
again, neglects 900\AA\ photons of non-stellar origin.
The upper limit on the escape fraction published by 
\cite{grimes07},
which compared only with the higher estimate of \flyc$_{,0}$ from
\citep{bergvall06}, 
remains unchanged at around 2\%.

{\em FUSE} data contains no spatial information about the LyC leakage. 
Both LyC and \lya\ are strongly affected by \hi\ 
covering but the nature of the effect is different for the respective photon
frequencies: LyC photons are absorbed but \lya\ photons are resonantly scattered, 
resulting in spatial redistribution. 
Since LyC photons do not resonantly scatter, their emission most likely traces their
production sites. 
The spatial redistribution of \lya\ can, to some extent, be mapped by our 
imaging technique. 
Only the central \lya\ emission region shows \lya/\halpha\ around case B level
and hence from knot \C, \lya\ is likely to be leaving the galaxy directly due 
to an outflowing neutral medium, or cleaned ionised holes. 
The fact that \lya\ may directly escape the galaxy from knot \C\
is suggestive that some LyC photons may also be doing so
and, in the limit that all the LyC radiation leaks from \C\, we obtain 
$f_\mathrm{esc} = 1/3$ for LyC at knot \C.
On the other hand, if the \lya\ emission from \C\ is purely due to the 
outflowing medium, this doesn't aid in the escape of ionising radiation and the
LyC photons must be leaking from elsewhere.
Without high-resolution LyC imaging, mapping of the \hi\ appears to be the only method 
by which location of the LyC leakage can be addressed.

\section{Conclusions}\label{sect:conc}

Using {\em HST/ACS} we have mapped, calibrated, and analysed the \lya\ emission
from nearby luminous blue compact galaxy Haro\,11. 
The \lya\ emission has been compared to: 
\halpha; 
\ebv\ as determined through both \halpha/\hbeta\ and UV continuum; 
the kinematic structure of the galaxy from previous studies; and 
the properties of the super star clusters. 
We have used archival X-ray data to map the hot outflowing gas and to put
constraints on the evolutionary state of the burst.
From SED fitting we have estimated and mapped the predicted Lyman-continuum
production at 900\AA, and compared this direct measurements.
Most notably: 
\begin{itemize}

\item{Our photometry reproduces spectroscopic fluxes as determined from the 
{\em IUE} satellite. 
The total \lya\ flux is found to be $79.6 \times 10^{-14}$~erg~s$^{-1}$~cm$^{-2}$, 
corresponding to a luminosity of $7.22 \times 10^{41}$~erg~s$^{-1}$.  }

\item{The escaping fraction of \lya\ photons is found to be $\sim 3$\%.}

\item{\lya\ shows almost no spatial correlation with \halpha. 
The \lya\ morphology shows a central bright emission region where \lya\ photons
escape the galaxy directly, surrounded by a low surface brightness diffuse halo
that results from multiple resonant scatterings. 
90\% of the \lya\ output is in the halo component.
\lya/\halpha\ at the values predicted by recombination theory 
are seen only in the brightest central regions.}

\item{Little correlation is seen between \lya\ and dust.
The main central \lya\ emitting region shows bright emission from a region
where \ebv\ is significantly greater than regions that show only \lya\
absorption. 
In the regions of halo emission, \ebv\ extends beyond 1 although \lya\ is
resonantly scattered and cannot feel the effect of such high extinction.
This again indicates that dust is not the major regulatory factor governing the 
\lya\ morphology, which appears to be driven more by the H{\sc i} distribution 
and its kinematic structure. }

\item{X-ray observations reveal a diffuse, soft component centred on the central
star forming knots, and showing hot wind-driven gas pushed outside the central starburst 
region as probed by UV bands.
The brightest X-ray regions are found to be spatially coincident with the
regions of highest surface brightness in \lya.
This indicates that \lya\ emission from central regions may be the result 
of a perturbed and outflowing ISM, accelerated by the release of
mechanical energy from the SSCs located around knot \C. }

\item{From the super star clusters, we find that peak \lya\ equivalent widths are 
small at the youngest ages (1~Myr), reach a maximum from clusters in the range 
2.5--4~Myr, and decline again to zero at ages $\gtrsim8$~Myr. 
This is qualitatively consistent with models of \lya\ emission resulting from
outflows in the neutral ISM.  }

\item{Haro\,11 is the only galaxy known to emit Lyman-continuum radiation
($\lambda < 912$\AA) in the nearby universe.
From fitting spectral synthesis models we estimate the continuum flux at
900\AA\ to be
$f_{900,0}=12.3\times10^{-14}$~erg~s$^{-1}$~cm$^{-2}$~\AA$^{-1}$.
This corresponds to an escape fraction of 9\% at 900\AA.  }

\end{itemize}

\section*{Acknowledgments}

MH and G\"O acknowledge the support of the Swedish National Space Board (SNSB)
and the Swedish Science Research Council (Vetenskapsr{\aa}det; VR). 
JMMH and EJB are supported by Spanish MEC grant AYA2004-08260-C03-03.
We thank Kambiz Fathi for the crash course in Voronoi tessellation.

\newpage
\appendix

\section[]{{\em Solar Blind Channel F140LP} aperture
	corrections}\label{sect:apcorr}

Synthetic PSFs were produced using five different power-law continuum 
slopes $(f_\lambda \propto \lambda^\beta)$ for 
$\beta \in [~-2,~-1,~0,~+1, ~+2~]$. 
The {\tt PHOT} task in {\tt IRAF} is used for aperture photometry.
Naturally, at large apertures there is little or no variation in the correction
as a function of $\beta$, although variations exist at the 0.014 magnitude level 
between $\beta=-2$ and $+2$. 

\begin{table}
\caption{{\em HST/ACS/SBC/F140LP} aperture corrections (in magnitudes) derived 
from {\tt TinyTim} at radii between 0.10 and 1.0\arcsec\ for various
power-law continuum slopes.}
 \centering
  \begin{tabular}{@{}cccccc@{}}
  \hline
  \hline
  Ap. radius      &    &    & $\beta$ &    &    \\
  $[$~\arcsec~$]$ & -2 & -1 & 0       & +1 & +2 \\
  \hline
  0.10 & 0.477 & 0.473 & 0.470 & 0.466 & 0.463 \\
  0.15 & 0.357 & 0.355 & 0.352 & 0.350 & 0.347 \\
  0.20 & 0.274 & 0.273 & 0.272 & 0.271 & 0.269 \\
  0.25 & 0.218 & 0.217 & 0.216 & 0.215 & 0.214 \\
  0.30 & 0.181 & 0.180 & 0.179 & 0.178 & 0.178 \\
  0.40 & 0.127 & 0.127 & 0.127 & 0.127 & 0.127 \\
  0.50 & 0.085 & 0.085 & 0.085 & 0.085 & 0.085 \\
  0.60 & 0.054 & 0.054 & 0.055 & 0.055 & 0.055 \\
  0.80 & 0.024 & 0.024 & 0.024 & 0.024 & 0.024 \\
  1.00 & 0.011 & 0.011 & 0.011 & 0.011 & 0.011 \\
  \hline
 \end{tabular}
\label{tab:apcorr}
\end{table}

\bsp

\label{lastpage}

\end{document}